\documentclass[acmtog,authorversion,screen]{acmart}

\AtBeginDocument{%
  \providecommand\BibTeX{{%
    \normalfont B\kern-0.5em{\scshape i\kern-0.25em b}\kern-0.8em\TeX}}}

\usepackage{xcolor}
\definecolor{red}{rgb}{0.95,0.4,0.4}
\definecolor{blue}{rgb}{0.4,0.4,0.95}

\definecolor{darkblue}{rgb}{0,0,0.8}
\definecolor{darkred}{rgb}{0.8,0,0}
\definecolor{darkgreen}{rgb}{0,0.5,0}
\definecolor{grey}{rgb}{0.6,0.6,0.6}
\definecolor{col1}{RGB}{232, 161, 148}
\definecolor{col2}{RGB}{148, 187, 232}

\usepackage{xspace}
\newcommand*{\eg}{e.g.\@\xspace}
\newcommand*{\ie}{i.e.\@\xspace}
\newcommand*{\etal}{et al.\@\xspace}
\newcommand*{\vs}{vs.\@\xspace}

\usepackage[ruled]{algorithm2e}
\usepackage{multicol}
\usepackage{enumitem}

\setcopyright{acmlicensed}
\acmJournal{TOG}
\acmYear{2022} \acmVolume{41} \acmNumber{3} \acmArticle{30} \acmMonth{3} \acmPrice{15.00}\acmDOI{10.1145/3506693}

\begin{document}

\title{LookOut! Interactive Camera Gimbal Controller for Filming Long Takes}

\author{Mohamed Sayed}
\email{mohamed.sayed.17@ucl.ac.uk}
\orcid{0000-0002-4074-3314}

\author{Robert Cinca}
\email{robert.cinca.14@ucl.ac.uk}
\orcid{0000-0003-4326-261X}

\author{Enrico Costanza}
\email{e.costanza@ucl.ac.uk}
\orcid{0000-0002-1842-8342}

\author{Gabriel Brostow}
\email{g.brostow@cs.ucl.ac.uk}
\orcid{0000-0001-8472-3828}

\affiliation{
  \institution{University College London}
  \country{United Kingdom}
}

\authorsaddresses{Wenbin Li created the inspirational re-localizing gimbal-controlled camera under UK EPSRC's EP/K023578/1. Mohamed is funded by a Microsoft Research PhD Scholarship. This work is supported by the UK EPSRC grant EP/R513143/1 for the UCL Interaction Centre (UCLIC). Our study was approved by the UCLIC Ethics Committee (UCLIC/1617/017).\\\\
Authors’ addresses: M. Sayed and G. Brostow, University College London, 169 Euston Road, London, NW1 2AE, United Kingdom; emails: mohamed.sayed.17@ucl.ac.uk,
g.brostow@cs.ucl.ac.uk; R. Cinca and E. Costanza, University College London, 66-
72 Gower Street, London, WC1E 6EA, United Kingdom; emails: {robert.cinca.14,
e.costanza}@ucl.ac.uk.}

\renewcommand{\shortauthors}{Sayed~\etal}

\begin{teaserfigure}
    \centering
    \includegraphics[width=\linewidth]{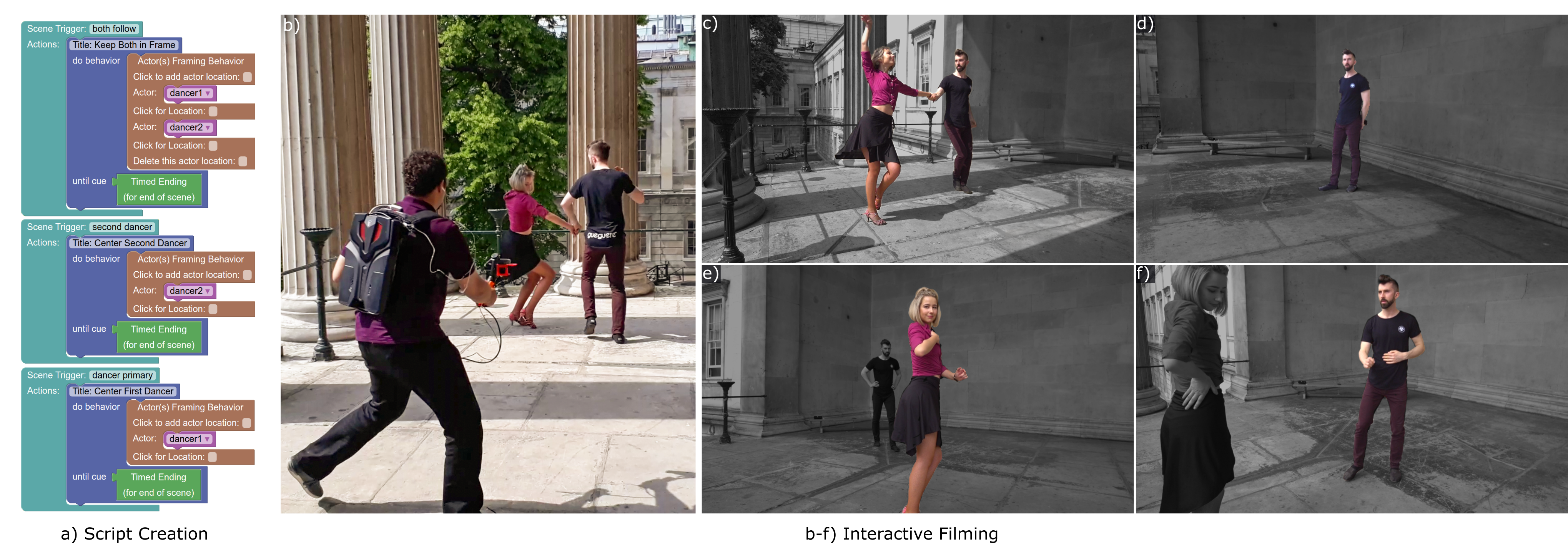}
    \caption{
    LookOut can take over the task of controlling where the camera is pointing when a camera operator is overwhelmed with other duties on the go, dynamically changing the camera's behavior based on where actors are, how a scene progresses, and what the camera operator instructs it to do. b) The user-worn LookOut rig consists of a light backpack computer, a hand-held motorized gimbal, dual cameras (normal and wide-view), earphones, a lapel microphone, and a joystick for initial setup. Before filming, the LookOut GUI a) enables a user to pre-script where the camera should point and its focal length. This involves creating camera behavior blocks that can be chained together to make scripts, callable during filming. A behavior can be as simple as a pan, or as complex as positioning multiple subjects in different parts of the frame, and they can be sequenced with scene specific cues. On boot, LookOut guides the operator, through text-to-speech, to enroll actor identities to its visual tracker, perform scene-specific initialization, and calibrate audio. c-f) Four frames from a LookOut-captured video, but with false-coloring to visualize which actor(s) LookOut is dynamically framing via its motorized gimbal to satisfy the operator's currently selected script. At the user's instruction, LookOut frames (c) both dancers, then (d) orients the gimbal to center on the male, then the female (e), and back to the male (f). The user receives audio feedback when switching between camera behaviors. Without a field monitor, the user can watch where they're going, while trusting our controller to handle their dynamic requests.
    }\label{fig:Teaser}
\end{teaserfigure}

\begin{abstract}
The job of a camera operator is challenging, and potentially dangerous, when filming long moving camera shots. Broadly, the operator must keep the actors in-frame while safely navigating around obstacles, and while fulfilling an artistic vision. We propose a unified hardware and software system that distributes some of the camera operator's burden, freeing them up to focus on safety and aesthetics during a take. Our real-time system provides a solo operator with end-to-end control, so they can balance on-set responsiveness to action \vs planned storyboards and framing, while looking where they're going. By default, we film without a field monitor.

Our LookOut system is built around a lightweight commodity camera gimbal mechanism, with heavy modifications to the controller, which would normally just provide active stabilization. Our control algorithm reacts to speech commands, video, and a pre-made script. Specifically, our automatic monitoring of the live video feed saves the operator from distractions. In pre-production, an artist uses our GUI to design a sequence of high-level camera ``behaviors.'' Those can be specific, based on a storyboard, or looser objectives, such as ``frame both actors.'' Then during filming, a machine-readable script, exported from the GUI, ties together with the sensor readings to drive the gimbal. To validate our algorithm, we compared tracking strategies, interfaces, and hardware protocols, and collected impressions from a) film-makers who used all aspects of our system, and b) film-makers who watched footage filmed using LookOut.
\end{abstract}



\maketitle

\section{Introduction}
Filming for journalism and movies is a creative and often collaborative process, where the budget dictates if the roles of director, director of photography (DP), and camera operators are fulfilled by a team, or rest on just one person's shoulders. Ultimately, the person holding the camera has the responsibility of delivering both the content and style that was agreed in advance, while safely adapting to dynamic changes on set.

After budget, time is the next biggest constraint. We consider two types of filming scenarios: one type where a journalist or documentary-maker must catch a one-off unrepeatable event, and the other type where actors and crew follow a storyboard with blocking, repeating the performance until the director is happy. Our system, called ``LookOut,'' is designed to help with both types of filming, if the aim is to capture a long take with a moving camera.

Long takes stand out as noteworthy and complex to choreograph in big-budget films\footnote{See the films \textit{1917}~\cite{Mendes1917} and \textit{Birdman~\cite{Birdman}}, both filmed to look like one take, versus Michael Bay's average shot length of 3 seconds~\cite{ShotLengthBlog}.}, though they are common for journalism, documentaries, and run \& gun videos - so the majority of working video/cinematographers. Moving the camera helps keep long takes interesting for the viewer~\cite{katz2004cinematic,brown2016cinematography,mascelli1976fivecs}. Steadicams~\cite{steadicam} and camera gimbals~\cite{saika2018camera, ursan2004camera} aid in filming these scenes by keeping the camera steady. Steadicams isolate translational motion through springs and arms and camera gimbals mainly isolate a camera from rotational movement of the carrier assembly by suspending the camera on a pivoted support with - often motorized - orthogonal axes. However, moving cameras and moving people stretch the attention of camera operators, who are trying to simultaneously walk about and adequately frame their stars. Usain Bolt was famously run over by a cameraman who suffered from task overload while steering a Segway at the World Athletics Championships in 2015. 

Speaking informally with independent film-makers, we found there was some interest in drone cinematography systems like \cite{Drone18,realtimeplanning,cineDrones}, 
but a strong desire for three things: 1) to have interactive control while filming, 2) a system that tracks indoors and outdoors without special costumes, and 3) ideally, to work with lightweight hand-held hardware, because drones are prohibited in many populated areas, and most countries require a pilot's license. This seeded our research process, which, with feedback and validation from filmmakers, has led to our proposed LookOut system (see Fig~\ref{fig:Teaser}).

The overall LookOut system serves as an interactive digital assistant for filming long takes with a camera gimbal. LookOut consists of software and 3D printed hardware that augments an existing lightweight motorized camera gimbal (\$130), with a video feed and rudimentary two-way speech-interface connected to a backpack computer. Without innovations, some of the individual components existed in principle, but would not integrate into a usable or responsive video-making algorithm. Therefore, our two main technical contributions are:
\begin{itemize}
\item A visual tracking system that detects and tracks actors robustly in realtime for extended periods of time, relying on a dynamic cost formulation for tracker/detection assignment, strategies for creating and maintaining a robust and space-efficient appearance history, and a recovery mechanism for minimizing distractions when reacquiring actors after occlusion.

\item A combined controller that dynamically balances script-induced constraints like smoothness and intentional framing to re-frame actors dynamically, while still being responsive to tracker outputs that have inherent noise and drop-outs.
\end{itemize}

The camera operator often wears many other hats, but from their perspective, during the critical moments of filming, the LookOut system responds to voice commands and follows alternative or sequential pre-specified behaviors. It rotates and stabilizes the camera within its joint limits, to follow the actors and to compensate for the operator's trajectory through the scene. For our experiments, operators didn't see a monitor while filming, so were free to look around and keep one hand spare as they walked, climbed, or cycled through different environments. 

\begin{figure}[!t]
  \centering
  \includegraphics[width=\linewidth]{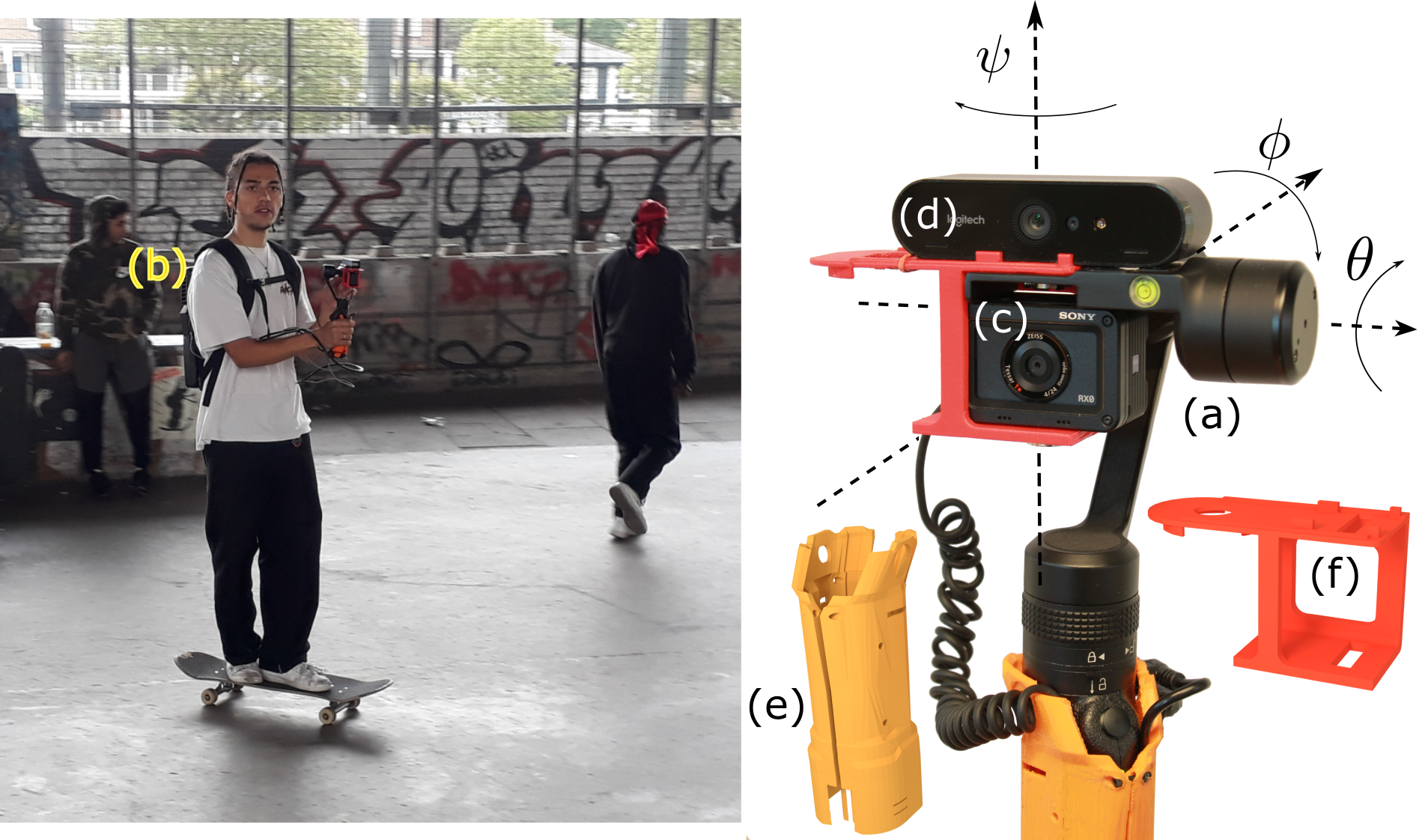}
  \vspace{0pt}
  \caption{{\bf A novice camera operator filming using the LookOut system:} (a) is an existing active camera gimbal, designed to stabilize mobile-phone filming. The mini-joystick is inactive by default, 
  The orange 3D-printed handle channels the cables and protects the USB connectors from being bumped. (b) is the backpack computer, connected to the gimbal by one USB cable and connected to (d) with another. Not shown, the backpack also has headphones and a lapel mic, for two-way speech communication with the operator. (c) is the primary ``star'' camera, recording high quality footage to local memory. (d) is the guide-camera, which has a wider field of view than (c), and whose video is fed to the backpack computer for real-time analysis. Star camera frame axes are represented with pitch ($\theta$), roll, ($\phi$), and yaw ($\psi$). The LookOut controller drives the orientation of the camera assembly. (e) Gimbal handle enclosure to allow for wire pass through and a comfortable grip. (f) Camera assembly engineered for balance, and alignment of camera optical axes}
  \vspace{-14pt}
  \label{fig:RigOverview}
\end{figure}

\section{Related Work}
The graphics community has a long history of exploring camera placement~\cite{christie2008camera} and control systems~\cite{GleicherWitkin92}, striving to be automatic and cinematic. For ``offline'' scenery special effects, motion control camera systems have been used since the work of computer graphics pioneer John Whitney in the 1950's~\cite{youngblood1970expanded}. While programmable camera trajectories can help with stop motion animation, and with layered compositing of scenery and special effects, they require hiring of specialized crew, are usually constrained to a short track, and the systems ignore actors and other dynamic events. We therefore focus this review on the context of our system, so following and framing of actors in video. This includes stabilizing gimbals, visual active tracking, and the efforts in drone cinematography.

\subsection{Steadicam, Stabilizing Gimbals, and Active Tracking}
Camera gimbals are essential for smooth video capture, especially when the whole assembly is held by a walking camera operator. 

Garrett Brown invented the Steadicam~\cite{steadicam} in 1975. The Steadicam allows for a camera operator to physically move the camera and simultaneously capture smooth footage. It has been famously used in many Hollywood film productions, including \textit{Rocky} (1976)~\cite{rocky}, \textit{Goodfellas} (1990)~\cite{goodfellas}, and \textit{Indiana Jones and the Temple of Doom} (1984)~\cite{indie}. Steadicams provide an extra layer of isolation from the camera operator compared to gimbals, in that they also dampen camera translation. Some are motorized to provide active stabilization and manual motorized control over the direction of the camera.  Although the camera operator no longer has to worry about keeping the camera steady, the operator must still point the camera while moving, either electronically through a joystick 
or manually by rotating the camera assembly. 

BaseCam Elecronics~\cite{basecam} develop different hardware and software components for the construction of stabilizing gimbals. Their firmware offers control and flexibility over every stabilization parameter. We build on top of their BaseCam Handy gimbal, which offers 3-axis control over camera orientation. Communication to the gimbal is achieved through a serial API that allows for online control and settings changes on the fly.

Many early active tracking systems focus on surveillance applications. Daniilidis~\etal~\cite{daniilidis}'s pan-tilt camera control orients a camera to focus on motion in a static scene. Dinh~\etal~\cite{anotherOne} and Funahasahi~\etal~\cite{faceptz} propose multi-camera or multi-focal length camera systems for identifying pedestrians through facial recognition. These systems are among the many that actively controlled pan, tilt, and zoom.

Closest to our own hardware is the DJI Osmo Mobile~\cite{djiOsmo}. It is a commercial real-time handheld active tracker. It uses a motorized gimbal and inertial measurement units (IMUs) to control a smartphone camera's orientation. The gimbal enables the user to create stabilized camera footage and select a single object to actively track. A smartphone is used as the camera and processing unit. The tracking algorithm is not made public. Unlike our system, users have no control over framing and complex scripting, and no ability to track multiple targets.

\subsection{Tracking} \label{trackerLit}
Generally, the ability of a tracker can be measured based on some high level performance criteria. Among them are speed, accuracy including robustness to ID switching - confusing another object with the target - or drift, number of trackable objects (usually one \vs many), robustness to appearance changes, and the ability to be run online. Most trackers in the literature are designed to some, but not all, of these. Our application requires robust tracking of a handful of targets for long durations (>20 minutes). Robustness to ID switches and target re-acquiring after occlusion, especially in busy and cluttered environments, are crucial to our use case since a target swap during filming would very likely ruin a take and cause delays. We focus almost entirely on trackers that can approach real-time speeds. 

The VOT challenges~\cite{VOT2017, VOT2018, VOT2019, Kristan2020a} cater to single target tracking of any class and includes benchmarks for RGBD and thermal trackers. The VOT Short-Term Challenge allows tracks to be reset, with a penalty and a timeout of five frames, to make use of the entire dataset. Trackers in the main VOT challenge are not required to deal with longer term occlusion and confidence reporting. In our use case, actors often appear and disappear as filming progresses. While the VOT Long-Term Challenge evaluates trackers with metrics that put a greater emphasis on longer term tracking (the average video is 2m04s long and contains 10 occlusions lasting 52 frames~\cite{VOT2019}). Other tracking datasets also contain long videos with occlusion ~\cite{Valmadre_2018_ECCV, moudgil2018long}; unfortunately, benchmarks on these datasets are either not maintained or trackers submitting not required to share implementation details. These benchmarks do not run trackers in a multiple object regime. 

A family of single object trackers are built on top of relatively lightweight Siamese network architectures~\cite{siam, CFnet, siammask, dasiamrpn}. Most notably, SiamMask~\cite{siammask} achieves state-of-the-art performance on the VOT2018 challenge at 50Hz. DaSiamRPN~\cite{dasiamrpn} includes a "distractor aware" module for reducing track loss errors after occlusion; it achieves first place on the VOT2018 real-time challenge, and second place on the VOT2018 long-term tracking challenge~\cite{lukevzivc2018now, VOT2018} at 110Hz. We experiment with both trackers and show how they are both prone to imposters of the same object type in long takes and cluttered environments.

The MOT challenge~\cite{MOT2016} provides performance metrics on trackers for multiple people in crowded scenes. The average shot length in MOT is \textasciitilde31 seconds with most targets exhibiting shorter life spans. While MOT includes metrics that measure ID swaps, resumed tracking with a new ID is still rewarded.

Among the lightweight high scoring MOT multi-person trackers, DeepSORT~\cite{deepSORT} and MOTDT~\cite{MOTDT} stand out. Both incorporate a tracking-by-detection paradigm and use a combination of IOU and appearance costs via ReID networks for assignment. Assuming detections are precomputed in advance, they could theoretically operate at 120Hz and 60Hz respectively. In Sec.~\ref{sec:Results} we compare against these trackers and show that while they are capable of tracking in dense scenes with short lived tracks, as in MOT, they are not robust to ID switches when tracking people in frame for longer videos, making them inadequate for our use case. 


While these trackers offer good performance across a wide range of metrics and for different classes of objects, no one tracker satisfies all the requirements of our use case, especially for people tracking.

\subsection{Automatic Drone Cinematography}
Though drones are contentious with safety restrictions in many countries, we share many objectives with drone-based cinematography. Skydio~\cite{skydio} and DJI~\cite{dji} provide multiple commercial drones with autonomous flying, self localization, and single actor tracking capabilities.

Drone cinematography is an active area of research. Ideas explored include actor pose driven drone flying~\cite{actiondrone}, controlling subject framing autonomously~\cite{quadcinedroneroberts, realtimeplanning}, constructing drone paths around user defined way-points~\cite{Drone18}, learning or mimicking shot style and kinematics from expert drone pilots~\cite{learnedcinedrone, fpvDrones} and using the Prose Storyboard Language (PSL)~\cite{psl} for actor framing and plotting drone paths~\cite{cineDrones}. Although these methods either use limited UIs and/or non-visual means of actor tracking (GPS and infrared markers), they showed promise for the concept of scripted and actor-driven camera control.

While we share the excitement around drone-based filming, drones are not always the correct, safe, or perhaps even legal tool for the task. Most actor driven shots take place in close quarters, with the camera closely following actors in the middle of the action. Further, while dubbed audio may be used in scenes, the noise they produce will ruin on-set audio.  

\subsection{Post-Filming Video Directing and Editing}

While this work proposes getting the right framing during filming, other work focuses on either fixing framing in post or automating some or all of the editing process~\cite{editingDavis2003}.

Many methods correct erratic and shaky camera movement in video~\cite{matsushita2006full, GrundmannKwatra2011, liu2013bundled, kopf2014first}. Grundmann~\etal~\cite{GrundmannKwatra2011} formulate L1-optimal camera paths in hand-held footage while incorporating framing constraints, notably a constraint on incorporating important features via a relevant saliency map, e.g. output from a face detector.

Gaddam~\etal~\cite{beVamsidhar2014} propose a system for both real-time and offline user controlled framing in high-resolution video. Grauman~\etal~\cite{makingGrauman2017} improve the state-of-the-art for automated $360^{\circ}$ to narrow field of view video editing by allowing for varied style, enabling zoom, and improving computational efficiency. Gandhi~\etal~\cite{multiVineet2014} propose a system for automatically extracting multiple clips from a single camera angle to assist editing.

Leake~\etal~\cite{DialogDrivenVideoEditingLeake2017} formulate a system for automatically editing together multiple takes of a dialogue driven scene with guidance on style taken as input from the user. ~\cite{editingDavis2003}. Wright~\etal~\cite{aiWright2020} describe and evaluate Ed, a system for automated camera and framing selection for live events. These methods are complementary to ours.

\section{LookOut System Overview} \label{interface}
At a very high level, the proposed LookOut system lets a user specify what they want to track, and then aims the camera gimbal at that target during filming. Achieving that aim required many iterations of hardware and software, user interfaces, and especially (1) innovations in long-term visual tracking and (2) a novel control system. Here, we outline the components of the system, and how they help the operator to design and safely film the long takes they want. 

A solo camera operator, without specialized programming skills, uses our GUI for offline pre-production, and our rig for live filming. We consider post-production only as part of Future Work. Interestingly, Leake~\etal~\cite{DialogDrivenVideoEditingLeake2017}, Wang~\etal~\cite{WYHYS-video-19}, and Zhang~\etal~\cite{ZhangRefocusableVideo2019} built interfaces that use learning to assist precisely with film-editing of existing clips. Instead, through our GUI, the user defines their intentions up-front - somewhat like telling an assistant what to expect. Those intentions are saved into scripts, that are later parsed by the LookOut control system during filming. On set, the camera operator wears a backpack-computer (see Fig~\ref{fig:RigOverview}) as the control-center and sensor-hub. The user also holds the camera gimbal in one hand, and has dialog with the LookOut controller, by wearing a microphone and headphone.

\subsection{High Level Components}
We give a brief overview of these components here, before providing their specifics in Section\ref{sec:trackerSection}, Section~\ref{controlSys}, and the supplemental material.

\textbf{GUI:} 
Before filming takes place, the camera operator uses LookOut's GUI to ``tell'' the camera gimbal how to behave and what to expect. The behaviors are chained together into a relative timeline. Instead of absolute times, user-specified cues will conclude and then trigger each subsequent behavior in turn. Through the script file saved by the GUI, non-programmer users instruct the LookOut control system with what to look for in the audio and video sensor inputs, and how to react. Please see the supplemental materials where we show the Blockly-based LookOut GUI for designing long takes. There, we explain how non-programmer users build script files by assembling chains of behaviors. A resulting script file encapsulates how one or more actors (and even non-actors) should be framed while filming. The script file switches between behaviors when triggered by user-controlled cues, that LookOut checks for continuously: Speech cues, Elapsed Time, Actor Appearance/Disappearance, Actor in Landing Zone, and Relative Actor Size. We are proud of the GUI for being easy to learn and for matching many of the wishes voiced by consulted film-makers.

\textbf{System Startup and Setup:}
When the LookOut hardware is first switched on, the user selects which scripts to load into the system. LookOut then parses these scripts and asks the user, through guided audio feedback, to enroll actors for tracking. The user adds an actor by pointing the camera roughly in the actor's general direction and pressing a button on a small joystick. LookOut guides the user for each additional actor. The system then prompts the user to utter each script-relevant speech trigger. This ensures all speech triggers are registered by LookOut using the user's current hardware audio configuration. 
LookOut informs the user that setup is complete and remains in \textit{Manual Mode} until the user requests \textit{Automatic Mode}.
Every mode switch and behavior trigger is met with audio feedback.

\begin{figure*}[t]
  \centering
  \includegraphics[width=\linewidth]{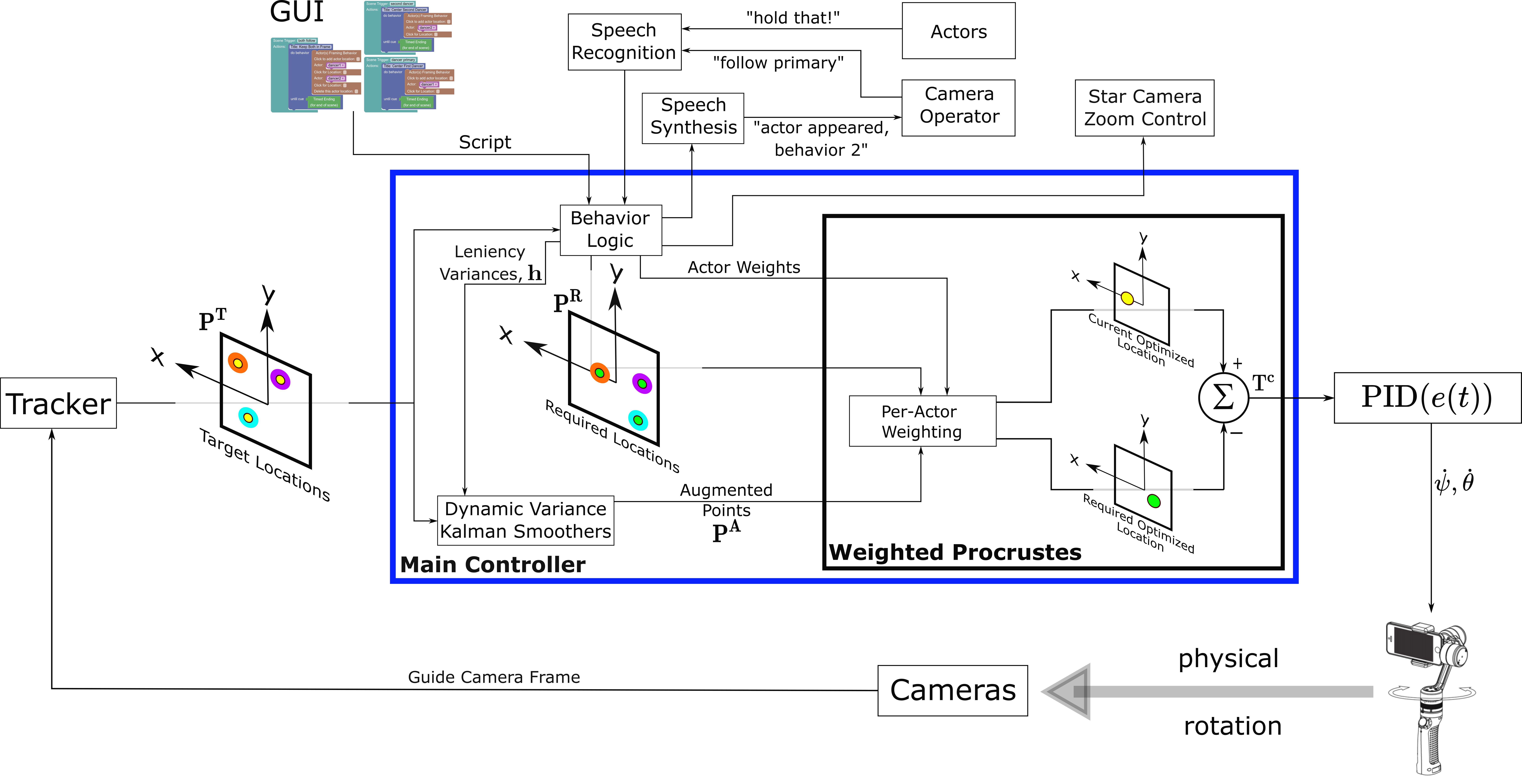}
  \caption{High level control loop view of how LookOut fulfills subject framing. On top, user inputs come in the form of the GUI during pre-production and through the use of speech commands on-set. At the bottom, the tracker converts guide camera footage into raw tracks, $\mathbf{P_T}$. All of these inputs enter the main controller (highlighted in blue and explained in Sec~\ref{controlSys}), whose job is to provide an error signal that will drive the gimbal through the PID~\cite{king2016process} controller. By modulating process variances, $\mathbf{h}$, the controller balances between responsiveness and smoothness for one or more actors. $\mathbf{h}$ is among the outputs from behavior logic, which had access to augmented track points from the last timestep (not pictured) and current target points, $\mathbf{P_T}$. $\mathbf{h}$ helps compute the augmented points, $\mathbf{P_A}$, which go into the weighted procrustes module (weighting explained in Sec~\ref{subsec:transitions}). The other main input to the procrustes module is the required locations for each actor, $\mathbf{P_R}$. Finally, the weighted difference between required locations and augmented locations drives the gimbal update. Not seen here is a velocity fading module that fades between different velocities at the transition from one type of behavior to another.}
  \label{fig:controlLoopFigure}
\end{figure*}

\textbf{Controller:} \label{sec:sceneUnderstanding}
The controller reconciles the input script(s) with incoming sensor data, to dynamically drive the gimbal motors. When a script sets out the camera behaviors, the controller listens for the relevant audio-cues, and analyzes the video feed to monitor spatial relationships between enrolled actors. It then dynamically drives the gimbal to achieve the desired framing and smoothness. Finally, it gives audio feedback to the user, so they know that the LookOut system is correctly following the script and the current actions. The control loop is described visually in Fig.~\ref{fig:controlLoopFigure}.

\textbf{Visual Tracking:}
Dynamic framing of one or more actors requires our system to follow along, monitoring where people are on-screen, even when they are briefly occluded or on the edge of the field of view (FoV). For these aims, we needed a visual tracker that can detect people and distinguish between them for long periods of time, despite imposter-objects, \eg people or things that could resemble the main actor(s). Our tracker balances the need for accuracy against the need to feed low-latency tracks to the controller.

\subsection{Hardware}
Here we describe the hardware and low-level software on which LookOut is built. Please see Figs~\ref{fig:RigOverview} for a close-up of hardware.

\textbf{Backpack:} Our system requires low latency feedback control in the wild. We use a VR backpack computer with a Quadcore Intel i7 7820HK CPU@2.90GHz 
and a mobile Nvidia GTX 1070 GPU.. The backpack can operate for 1.5-2 hours, allowing for very long shots and multiple takes, and is light at 3.6kgs.

\textbf{Stabilizing Gimbal:} We use the Basecam Handy gimbal 
to carry the camera assembly. The gimbal is programmable through a serial API and allows high speed low latency control and telemetry data transfer up to 80Hz. The gimbal has an Inertial Measurement Unit (IMU) on the camera frame assembly, and an encoder for each axis for tight closed loop feedback control. We have exclusive control over velocities on yaw (${\psi}$), pitch (${\theta}$), and roll (${\phi}$) on the camera frame assembly, regardless of the orientation of the handle.  We disable any internal low pass filters on velocity to ensure controllability. We tune the gimbal's internal proportional integral derivative (PID)~\cite{king2016process} loop for the tightest possible axis velocity control, while ensuring loop stability, given our camera array.

\textbf{Camera:}
We use two cameras in our system. One serves as a guide camera for visual tracking over a $90^{\circ}$ field of view. It operates at 60Hz and at a resolution of 1280$\times$720. We decouple roles a camera must perform by using a separate camera for capturing high quality footage - which we call a \textit{star camera}. This configuration was preferred by filmmakers in our initial scoping. It allows for cinematic freedom over camera parameters used for filming, without sacrificing preferred parameters and hurting the performance of the visual tracking pipeline. We design and 3D print a carrier assembly for the cameras, shown in Fig~\ref{fig:RigOverview}. It maximizes the balance on all gimbal axes, while minimizing the distance between the optical centers of both cameras within the gimbal's confined space. 

\textbf{Remote Screen:}
We use a remote HDMI transmitter and screen when turning the system on. Once the system is setup, the screen is put away. 

\textbf{Audio:}
The user wears a lapel mic and earphones to speak commands to the system during filming, and to receive feedback throughout actor-enrollment and filming. We use an online wake word detection framework, Porcupine~\cite{porcupine}, for recognizing speech commands.

\section{Tracker} \label{sec:trackerSection}

To achieve LookOut's aim of framing actors, the system needs to know their locations in screen space. The tracking component must work reliably for filming impromptu run \& gun situations. Attaching real tags to actors such as in~\cite{realtimeplanning, cineDrones} is often impractical. To this end, the tracker must be completely visual in nature. 
The requirements of the tracker are that it must:

\begin{enumerate}
    \item be capable of locating multiple targets of interest simultaneously, with a focus on actors,
    \item reacquire actors when they appear back in frame, while being robust to ID switches, and
    \item maintain a high online refresh rate (>30Hz) and low latency to ensure fast actor movements are captured and acted on by the control feedback loop discussed in Sec~\ref{controlSys}. 
\end{enumerate}

We cover the current state-of-the-art in Sec~\ref{trackerLit}. Broadly, the trackers that are fast enough (>20Hz) fall into two categories, single object trackers aimed at the VOT~\cite{VOTmetrics} and OTB~\cite{OTB2013} challenges, and multi-target trackers from the MOT~\cite{MOT2016} challenge. We compare against the best trackers from these challenges in Sec~\ref{subsec:trackerResults}. Notably, while single object trackers like DaSiamRPN~\cite{dasiamrpn} and SiamMask~\cite{siammask} perform well when keeping track of an object in frame, they are prone to tracking imposters when an object is occluded and reappears in frame, not satisfying (2). To satisfy (1), a different instance of each tracker would need to run separately for each actor; this compromises (3) since the runtime now scales linearly with the number of actors. 

For trackers competing in the MOT~\cite{MOT2016}, almost all trackers use tracking-by-detection; these trackers suffer a relatively small penalty for each additional target, but require a real-time detector with a good compromise of accuracy and speed\footnote{MOT trackers take detection bounding boxes for granted in the benchmark.}. However, the MOT benchmark is run on scenes whose mean length is only $\sim$31 seconds, where targets only occasionally change view throughout their short life, and rarely reappear after long term occlusion with a small penalty given for ID switching. We take inspiration from high scoring trackers in the MOT benchmark, DeepSORT~\cite{deepSORT} and MOTDT~\cite{MOTDT}, but add three contributions: 
\begin{itemize}
    \item a reworked cost structure for detection/track assignment, with a concentration on tracking a handful of targets robustly,
    \item a recovery phase and mechanism, and
    \item a set of lightweight long term appearance-encoding history-management strategies.
    
\end{itemize}

Our tracker relies on an appearance-encoding history for differentiating actors and other people during filming. A reliable per-actor appearance-encoding gallery is important for tracking and recovery. All three components, explained below and in pseudocode in the Supplementary Material, 
focus on maintaining correct IDs for each actor, especially after occlusion.

\textbf{Cost Formulation and Data Association:} 
Our tracker minimizes the the cost of assigning targets $T = {t_1, ..., t_i}$, including appearance and bounding box information, to a set of detections in the current frame $D = {d_1, ..., d_j}$. Taking inspiration from DeepSORT~\cite{deepSORT}, we combine ${c^{\text{IOU}}_{ij}}$, the IOU bounding box cost~\cite{jaccardIndex, MOT2016}, with $c^{\text{f}}_{ij}$, the cosine distance on appearance features, derived from the Siamese network in~\cite{zheng2016mars}. We do not use a Kalman filter state based cost, as detections from our choice of lightweight detector, tiny-YOLOv3, are very noisy spatially over time. 

IOU costs are useful when a target is in isolation, but useless when overlaps occur or when coming out of a long occlusion. Appearance costs on the other hand are crucial for re-identifying the target after long occlusion, but a collection of appearance features, capturing the appearance of the target under different lighting and self occlusion, must be accumulated before they can be relied on. To this end, we formulate a dynamic cost structure specific to each target, that emphasizes robustness by relying on IOU when no more than one detection competes for the same target, and the appearance cost when a target is crowded. Nominally, the cost for associating a particular target and detection, $c(t_i, d_j)$, is 

\begin{center}
\begin{equation}
c(t_i, d_j) =
  \begin{cases}
               c^{IOU}_{ij} & \text{if $c^{IOU}_{ik} > \tau^{\text{overlap}}$ where $k \neq j$ } \\
               c^{\text{feature}}_{ij} + c^{IOU}_{ij} & otherwise. \\
  \end{cases}
\end{equation}
\end{center}

${\tau^{\text{overlap}}}$ is the cost of assigning the track $i$ to another detection $k$ and is set to a high strict value to prevent ID switches when a target is occluded by other people (other $k$s). To further reduce target switches, a track/detection pair are deemed incompatible if either the IOU cost or the appearance cost exceed defined low maximums.

We assign $c^{\text{f}}_{ij}$ the cost of the lowest cost match between a target's appearance encodings and that of a detection $d_j$. Although we take measures to exclude rogue imposter encodings, a single matched feature encoding can produce an incorrect match. To mitigate this, we take an average of the $N$ lowest appearance costs from the target's history and we disallow a match between this combination of track and detection if it exceeds a predefined maximum.

Finally, all costs are passed along to a linear assignment step~\cite{kuhn1955hungarian} where globally optimal target and detection assignments are found.

\textbf{Recovery:} Actors of interest will go into planned or unplanned short and long term occlusion throughout filming. During occlusion, the tracker must not confuse imposters with actors, and should then recover these actors when out of occlusion. We use appearance costs, $c^{\text{f}}_{ij}$, exclusively for this step. However, appearance encodings are temporally noisy, so an imposter detection might present a noisy appearance encoding in one frame that matches to a lost target. To prevent these types of false matches, we define a recovery phase that is begun when a detection is matched to a lost target. For a target to come out of recovery, it must be matched to a detection for $R$ sequential timesteps, where $R$ is decided dynamically. This mechanism sacrifices a few frames of tracking for recovery in the short term, but greatly improves the tracker's long term tracking ability and its resistance to ID switching. We test our tracker without a recovery step in Table~\ref{tab:trackerAblationResults}.

\textbf{Feature History Management:} In dense scenes and in a target's recovery phase, the tracker relies solely on each target's appearance encoding gallery, $\mathcal{R}_i=\{r_1, ..., r_L\}$ for data association. Ideally, an infinitely sized history would allow for the most accurate representation of the target's appearance. However, encoding comparisons for calculating appearance costs would get expensive with longer target life cycles - 10 minutes at 30Hz yields 18,000 appearance encodings. A common solution~\cite{deepSORT, MOTDT} is to restrict the gallery to the last $L$ encodings. This strategy works well for short track life in short sequences as in the MOT challenge; however, this is less successful for longer sequences where a target may reappear either with different lighting or pose than when they went into occlusion. Table~\ref{tab:trackerAblationResults} shows the performance of a tracker with a naive last-$L_k$ encodings history. We address the rapid increase in the gallery's size by selectively adding encodings to the appearance gallery on every time step. An encoding is added only if it is sufficiently distant, via the cosine distance, from all other encodings in the gallery. This slows down the growth of the gallery by an order of magnitude and prioritizes space and time on informative encodings.

When a target is crowded by many detections, encodings produced with occluded bounding boxes might later allow an impostor to match the target incorrectly. To address this, encodings are added exclusively in normal tracking - when only one detection competes spatially for the current target. In Table~\ref{tab:trackerAblationResults}, a tracker without this check is referred to as Faulty Encodings.

Although these steps help reduce the expansion of the gallery's size and maintain its integrity, they only delay the pruning problem when the gallery is full. Informed techniques that cluster encodings to select the most informative encodings are iterative and time consuming in this high dimensional space - k-means consumes 7ms for each target. Alternatively, a simple and effective solution is to randomly sample $L_k$ from the gallery when it is 10\% larger than $L_k$. This has the effect of maintaining new appearances of a target while keeping a fading memory of older appearances for longer, since with every sampling step, encodings of an older age stamp are less likely to be propagated forward.

\textbf{Speed:} As mentioned previously, MOT provides detection for granted and trackers do not report detection time. A survey of the detection field shows that single shot object detectors~\cite{YOLOv3, ssd}, are best suited for their trade-off of speed and performance. We use people, cars, and bicycles detections from tiny-YOLOv3~\cite{YOLOv3} in our system. We tune the Kalman Filters used for tracking updates to reduce temporally noisy detections from tiny-YOLOv3 before being passed to any control loops down the pipeline.

\textbf{Subject Enrollment:} Our tracker requires one frame to enroll an actor and can track subjects immediately. An extra step can be taken to build up an initial appearance history by having the subject turn around and ideally walk once through the scene.

As shown in the Supplementary Videos, we also experimented with DaSiamRPN~\cite{dasiamrpn} which allows enrollment of novel objects, such as a shop window and a garden gnome.

\section{Control system for framing actors} \label{controlSys}

We drive the camera orientation to re-frame actors dynamically over time. The controller reconciles live tracker data with the user's instructions and then drives motors on the gimbal to adjust the camera assembly's orientation to achieve the user's desired framing of one or more actors. The interface for user instructions is discussed in the supplementary materials, and actor location tracking was in Section~\ref{sec:trackerSection}.

The visual servoing community has made tremendous progress in constructing methods for moving cameras and robotic arms to desired positions in space and/or orienting them based on some external visual signal~\cite{visualServoingSurvey}. The bulk of visual servoing use-cases are in robot end effector control in manufacturing. Usually these methods involve the solution of a Jacobian matrix~\cite{highperfServ, newApproachVisualServoing} that encodes tasks and joint movement constraints. While some work explores modulating the variance of the mean position of all visual points of interest in image space~\cite{modulatingTasksVariance, quadRotorServoing}, none has provided a transparent formulation for controlling per target variance nor does one provide a framework for gradual change between different tasks and constraints. We borrow themes from the visual servoing literature while constructing a task specific control scheme.

Appealing camera positioning and orientation is essential for effective video game design, as such the video gaming has generated methods and implementation tricks for implementing dynamic cameras that follow in-game action on-the-fly~\cite{gameCameras}. While these methods assume that targets are known with certainty and that control over camera properties is instantaneous, we take hints from the community when designing our own control scheme and incorporate strategies to both mitigate and cope with real world noise.

At a high level, the controller is a closed loop feedback system with proportional-integral-derivative(PID)~\cite{king2016process} controllers that minimize an error signal, $e(t)$, by modifying the camera frame's yaw and pitch over time. $e(t)$ is an abstraction of the error between real-time dynamic actor locations and desired user framing encapsulated in the script. If we simplify camera space conversions, ignore noise, and assume only a single tracked target, then $e(t)$ is just the screen space difference between the actor's tracker location and the user's screen space requirement for actor framing, with both $x$ and $y$ components. Errors in $x$ and $y$ are corrected by changing the camera frame's yaw ($\dot{\psi}$) and pitch ($\dot{\theta}$) respectively. The corrections are handled by PID controllers, so

\begin{equation}
\dot{\psi} = \text{PID}(e_x(t)) \text{ and }
\dot{\theta} = \text{PID}(e_y(t)).
\end{equation}

We tune our PID controllers using a relaxed version of the Ziegler-Nichols procedure~\cite{pidopt} to achieve the tightest response possible, while minimizing overshoot, given delay and processing constraints. Note that these are camera frame radial velocities and not direct motor torque commands. The underlying gimbal camera assembly radial velocity stabilization is tuned in the gimbal's firmware and is not discussed here.

This abstracted version of e(t) is suitable for a single actor and will produce erratic camera motion since raw tracker locations are noisy either due to tracker inaccuracy or due to subtle actor movements. This is fine if the preferred style is very erratic unnerving camera motion with a random component due to noise, but not for any other desired style. Tuning the PID controllers to be lazy would ignore noise and allow for a lazy camera, but would erode control over all actor driven camera behavior and remove responsiveness when responsive corrections are required. Other design considerations include handling behavior transitions and tracker dropouts, where potential camera jerks are likely and a single one would ruin a take. The controller must handle a variety of filming scenarios and behaviors - from single actor to multi actor, from action scenes to calmer slower paced scenes, and the transitions between them. We therefore design the controller to pursue these design objectives:

\begin{enumerate}
    \item \textbf{Achieve desired user framing:} on every loop, the system should minimize the difference in required actor framing \vs actual actor framing. Logical compromises should occur when framing multiple actors at once.
    \item \textbf{Only move the camera if motivated:} the user can provide an ellipse for each actor in each behavior, indicating an area around the actor. There, their movements do not result in camera frame reorientation. The controller should also ignore noise from raw tracker estimates so the camera does not oscillate and produce unpleasant motion. This is discussed in Section~\ref{subsec:leniency}.
    \item \textbf{Enable smooth transitions:} As behaviors change, different actors come in and out of scene. The transitions between different actors must be smooth. This is discussed in Sec~\ref{subsec:transitions}.
\end{enumerate}

The $e(t)$ signal driving the PID controller is computed based on these objectives. Specifically, $e(t)$ comes from a weighted Procrustes module, that we simplified: it aligns the current $2$D actor location(s) with the location(s) required by the user, subject to the available degrees of freedom. We found that in-plane rotation for framing wasn't helpful. Therefore, in all our experiments, we used a Procrustes model that simply finds the translation vector $\mathbf{T^c}$ as the weighted difference of the average actual locations and the average required location; $\mathbf{T^c}$ acts as our error vector $e(t)$. We also extend this to account for star camera zoom by scaling the input points appropriately. To address (2), we add "Leniency," where instead of passing raw tracker components to compute $\mathbf{T^c}$, we instead produce dynamically decoupled and filtered Augmented locations in Sec.~\ref{subsec:leniency}. To address (3), we modify the influence each actor has in current framing given transitions and tracker confidence and introduce filtering on required script behavior in Sec.~\ref{subsec:transitions}. Fig.~\ref{fig:controlLoopFigure} provides a high level overview of the control system's components.

\begin{figure}[ht]
  \centering
  \includegraphics[width=\linewidth]{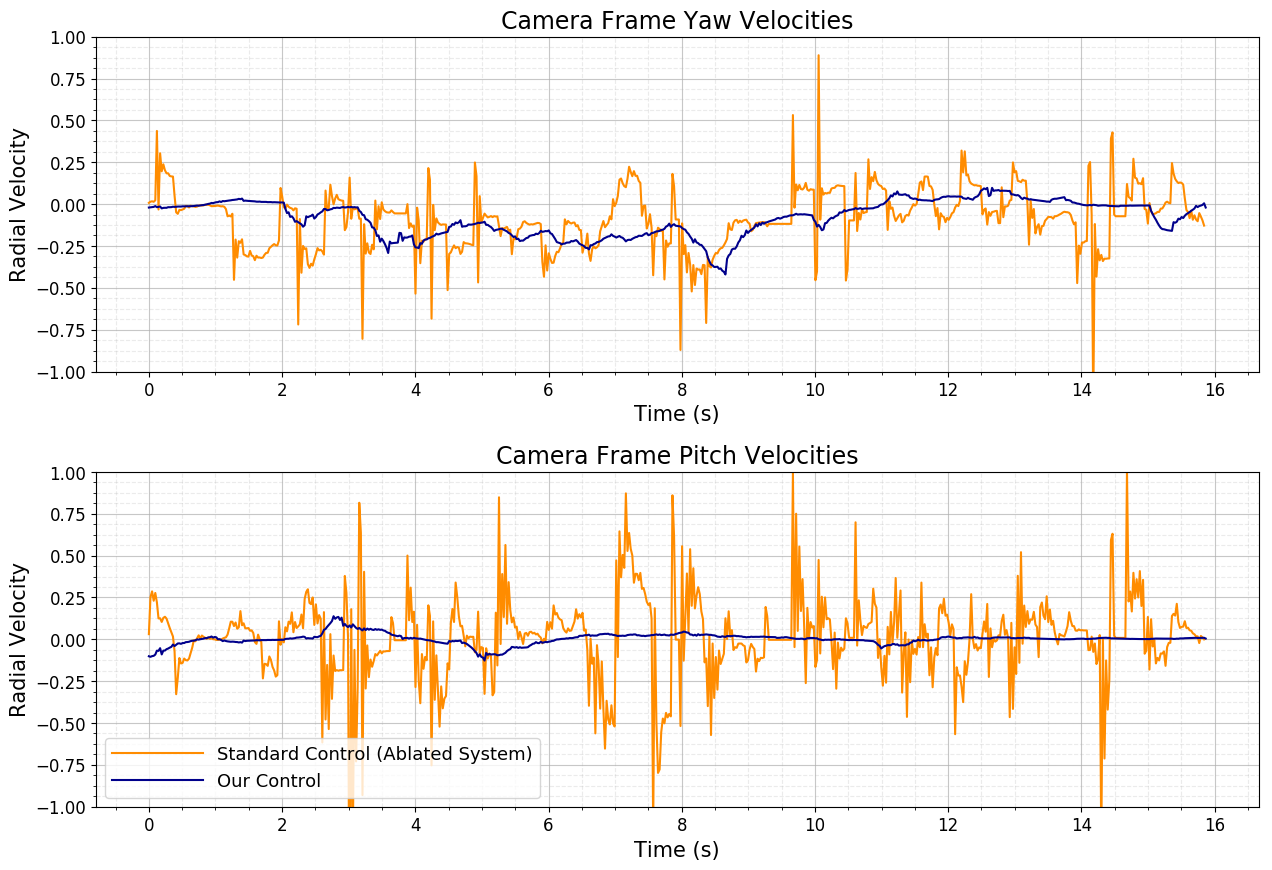}
  \caption{Yaw (top) and pitch (bottom) camera frame radial velocities for both our full system and ablated control throughout the control ablation running scene. The standard controller (ablated system) does not ignore tracker noise and translates changes in perceived actor location from the tracker directly to an error in the PID controller. This leads to massive over corrections and an uncontrollably erratic camera. Instead, our full controller can handle tracker noise and camera translation by using modifiable leniency (Sec.~\ref{subsec:leniency}) on raw tracker locations and applying control weight adjustment (Sec.~\ref{subsec:transitions}). Note that these are not gimbal motor velocities, rather these are target radial velocities for the camera frame to achieve. Raw gimbal axes velocities and torque are a function of required camera frame velocities and external forces acting on the gimbal assembly.}
  \label{fig:controllerAblation}
\end{figure}

\subsection{User Defined Motivated Camera Movement} \label{subsec:leniency}

Minimizing the difference between actor locations given by the tracker, $\mathbf{P^T} = \{\mathbf{{p^{T}_1}}, ..., \mathbf{{p^{T}_n}}\}$, and required actor locations, $\mathbf{P^R} = \{\mathbf{{p^{R}_1}}, ..., \mathbf{{p^{R}_n}}\}$, using Procrustes satisfies (1). However this wouldn't filter noise, either from the tracker or abrupt camera translation, and wouldn't allow for selectively ignoring small actor movement. Instead, to satisfy (2), we define \emph{tracked-smoothed-augmented} points (``Augmented'') $\mathbf{P^A}$ that are smoothed versions of $\mathbf{P^T}$ and use those to compute $\mathbf{T^c}$. The obvious means of making augmented versions $\mathbf{P^A}$ is via Kalman filtering, so 

\begin{equation}
\mathbf{{p^{A}_i}} = \text{KalmanFilter}(\mathbf{p^{T}_i}, \mathbf{h_i}),
\end{equation}

where $\mathbf{h_i}$ is a process variance. A high $\mathbf{h_i}$ means an augmented point follows its tracked point quickly, allowing for an immediate change in the error term for that actor and an immediate correction signal from Procrustes resulting in a very responsive camera to actor movement. A small $\mathbf{h_i}$ allows for the opposite: each $\mathbf{p^{A}_i}$ lazily follows its track point $\mathbf{p^{T}_i}$ resulting in a less immediate corrective signal and less eager camera panning.

However, fixing $\mathbf{h_i}$ limits user control. Ideally, there should be definable areas of forgiveness around an actor where small movements are ignored. Setting a small $\mathbf{h_i}$ allows this, but this would ignore actor movements outside of this area when they do matter. Instead we modulate each $\mathbf{h_i}$ based on $\mathbf{d^{LE}_{i}}$, the current discrepancy between a tracked point $\mathbf{p^T}$ an its augmented point $\mathbf{p^A}$ from the previous time step. We make $\mathbf{h_i}$ proportional to $\mathbf{d^{LE}_{i}}$, so that with a small $\mathbf{h_i}$ the Kalman filter will ignore new updates given by $\mathbf{p^{T}_i}$  and instead choose to maintain the older location of $\mathbf{p^{A}_i}$. As a point $\mathbf{p^{T}_i}$ moves too far from its $\mathbf{p^{A}_i}$, the distance, $\mathbf{d^{LE}_{i}}$, ramps up $\mathbf{h_i}$ and the Kalman filter is more sensitive to new incoming updates via $\mathbf{p^{T}_i}$, so $\mathbf{p^{A}_i}$ follows $\mathbf{p^{T}_i}$ more closely.

The relationship between $\mathbf{d^{LE}_{i}}$ and $\mathbf{h_i}$ is user definable and based on a family of exponential functions. We define a set of aesthetic parameters for each axis:

\begin{itemize}
    \item Zero Error Lift, $v$: This forces a non-zero value when $\mathbf{d^{LE}_{i}}$ is at zero. The result of a high $v$ is an immediate responsive pan from the camera when the actor moves small distances from rest.
    \item Agnostic Gap, $a$: This defines how much distance the actor has to travel before the camera pans,
    \item and Curve Profile, $q$: This defines the ramp up at the edge of the allowed area of leniency and determines how sharply the camera will pan when an actor begins to leave that leniency area.
\end{itemize}

We also set a hard limit on $\mathbf{h}$ via $\mathbf{\eta}$. This cap limits the impact from temporal instabilities in the tracker, and was experimentally set to 0.01 for vertical motion and 0.05 on horizontal motion in all experiments. We include a qualitative experiment for demonstrating these smoothing functions and raw tracker values in the supplemental videos. For each actor, each component of $\mathbf{h}=(h_x, h_y)$ is computed as:

\begin{equation*}
\begin{aligned}
h_x &= {\eta}_x\,\text{clamp}(0, 1, e^{q_x(d^{LE}_x-a_x)}+ v_x) \text{ and} \\
h_y &= {\eta}_y\,\text{clamp}(0, 1, e^{q_y(d^{LE}_y-a_y)} + v_y).
\end{aligned}
\end{equation*}

\begin{figure}[h]
  \centering
  \includegraphics[width=\linewidth]{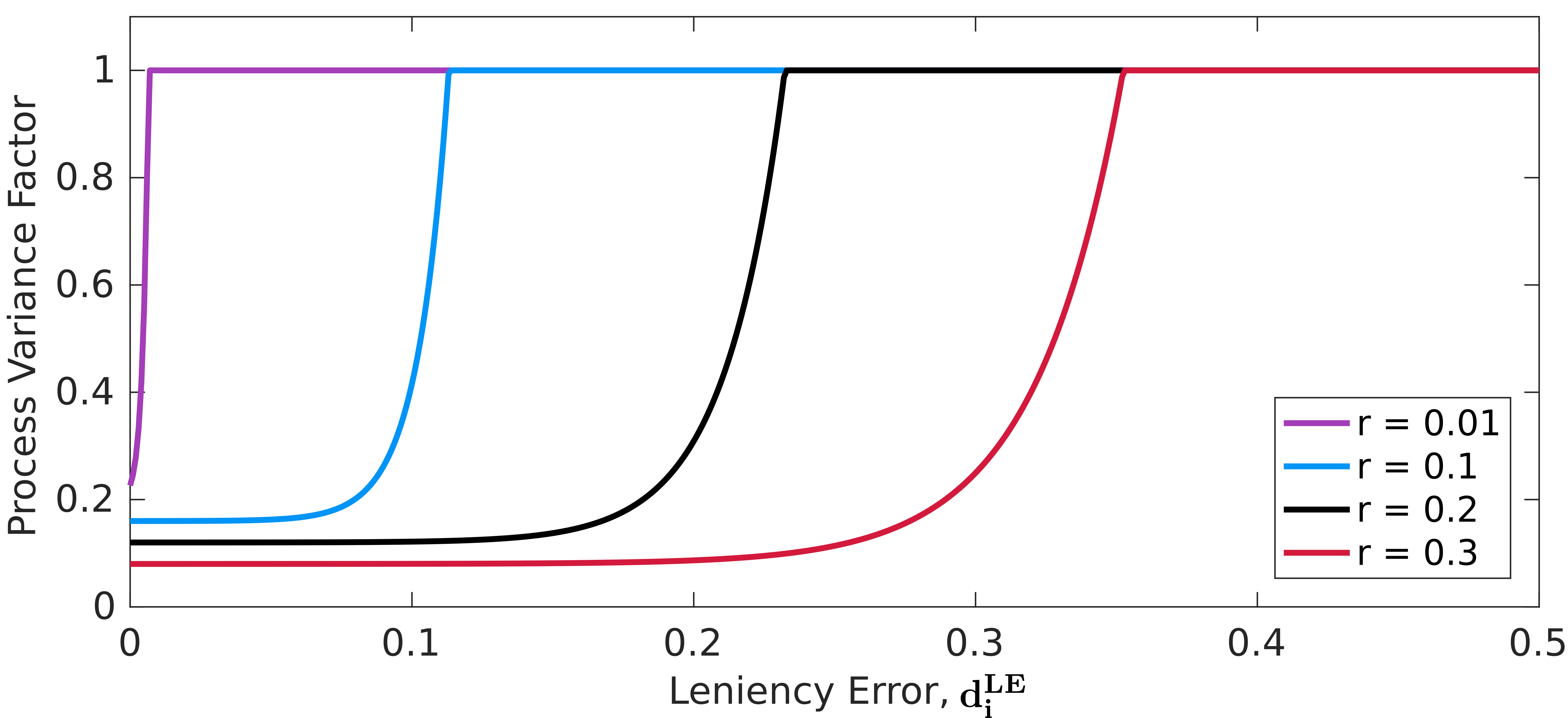}
  \caption{Curve profiles for different user prescribed leniency radii. These radii represent areas around the actor where where camera panning is attenuated if the actor moves in that area. The y-axis is applied to $\eta$ to produce each Kalman filter's process variance, $\mathbf{h}$. The x-axis is the difference, $\mathbf{d^{LE}_{i}}$, between the augmented version of the actor's location from the previous timestep, $\mathbf{{p^{A}_i}}$, and the raw tracker location, $\mathbf{{p^{T}_i}}$, and is normalized relative to screen space size. A smaller ellipse radius limits the area where the actor can move without a camera pan, as the process variance ramps up immediately. A larger ellipse allows for more actor movement before the camera starts panning.}
  \label{fig:leniencycurve}
\end{figure}

These equations are not obvious, but the three input parameters have interpretable connections to the radii $(r_x, r_y)$ of each ellipse drawn by the user in the GUI. The functions relating radii $\mathbf{r}$ to each of these parameters are given in the Supplementary Material. 
In brief and for a single axis, for a large $r$, $v$ is reduced such that almost no movement occurs at zero error, $a$ is made to satisfy the distance defined by $r$, and $q$ is set so that the transition is smooth. Conversely, for a small $r$, $v$ is kept high for immediate reaction, $a$ is set so that the point at which the curve increases happens earlier, and $q$ is set so that the curve is sharp. See Fig~\ref{fig:leniencycurve} for different curves corresponding to different user input radii. We include an example of multiple actor leniency in the supplemental video.

\begin{figure}[h]
  \centering
  \includegraphics[width=\linewidth]{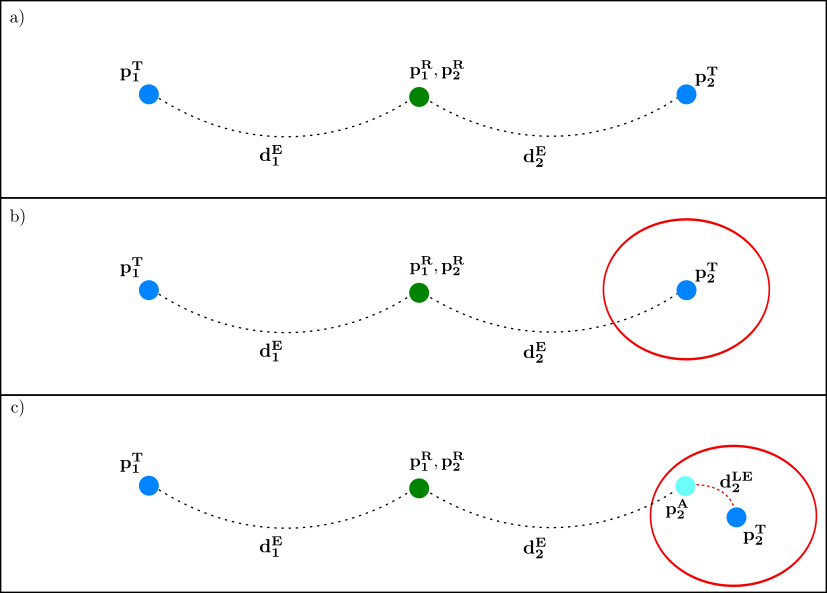}
  \caption{a) b) and c) show an example where the user specifies that both actors should be framed in the center given by required locations $\mathbf{p^{R}_1}$ and $\mathbf{p^{R}_2}$. However, the actors' relative locations at $\mathbf{p^{T}_1}$ and $\mathbf{p^{T}_2}$ make it impossible for that requirement to be fulfilled. As such, the best framing possible at steady state is where both are equidistant from the center. In a) no leniency is defined, and so a movement by either $\mathbf{p^{T}_1}$ or $\mathbf{p^{T}_2}$ will need a new optimization and the camera pans. In b) and c) leniency is required on actor $2$ given by the red ellipse defined by the user, i.e. if the actor at $\mathbf{p^{T}_2}$ moves within the ellipse, the camera should not respond. b) a naive solution to achieve leniency is to attenuate the error term $\mathbf{d^E_2}$ when the target $\mathbf{p^{T}_2}$ is close to the point of the optimization at steady state (where $\mathbf{p^{T}_2}$ sits). However, since this is a less than ideal framing with both required points at the center, a new optimization will be found that improves $\mathbf{d^E_1}$ and the camera pans, disregarding leniency. Instead in c) we formulate a new augmented point $\mathbf{{p^{A}_2}}$ that is output from a Kalman filter on $\mathbf{p^{T}_2}$ whose process variance is modulated by the distance $\mathbf{d^{LE}_2}$ w.r.t to the ellipse. The actor and ellipse at $\mathbf{p^{T}_2}$ can move around the augmented point, $\mathbf{{p^{A}_2}}$, and as long as the augmented point is in the ellipse, the process variance $\mathbf{h_2}$ remains low and the augmented point at $\mathbf{{p^{A}_2}}$ doesn't move with $\mathbf{p^{T}_2}$; the error term $\mathbf{d^E_2}$ remains low since $\mathbf{{p^{A}_2}}$ remains stationary although the actor at $\mathbf{p^{T}_2}$ has moved, and so the camera does not pan to compensate. When the actor leaves this ellipse, $\mathbf{h_2}$ is ramped up, $\mathbf{{p^{A}_2}}$ moves to follow $\mathbf{p^{T}_2}$, the error term $\mathbf{d^E_2}$ changes, a new optimized framing is found, and the camera pans.}
  \label{fig:leniencyFailureFigure}
\end{figure}

Note that the naive solution of simply weighting the error associated with the $i$th actor to zero when the actor is in some allowed radius will not achieve multi-actor leniency. Most situations result in a sub-optimal optimization where required actor locations are not fulfilled perfectly due to physical limitations. A zero weight for an actor would result in a new optimization and, counterintuitively, produce camera motion when none was required. See Fig.~\ref{fig:leniencyFailureFigure} for an illustration of this.

\subsection{Actor Transitions and Path Behavior} \label{subsec:transitions}
To allow for smooth transitions between subjects, satisfying (3), each actor is assigned a weight, $w_i$, that modifies the actor's error term in the Procrustes optimization. When an actor appears in frame and is part of the current behavior, their weight is increased progressively. Their weight is decreased when they either disappear from frame, because of occlusion or tracking failure, or are no longer included in the behavior. 

For each actor, we also apply Kalman filters on user selected required points as they transition between different behaviors so that no discontinuities occur. Separately, a behavior can be intentionally shaky to give the viewer a hand-held impression. To achieve shakiness or intentional banking behavior (like an airplane changing course), the controller reads the gimbal IMU accelerations on the camera's horizontal axis, applies smoothing, and actuates the roll axis. See the supplemental video for an example of a path behavior.

\subsection{Focal Length Control} \label{subsec:zoom}
We make star camera focal length control available to the user in a few ways. Standard operation modifies the set of required points, $\mathbf{P^R}$, to match star camera framing at the current zoom level. We do this by applying a scaling matrix with knowledge of the camera intrinsics at each zoom interval. The new required points used as input to the optimization are now

\begin{equation*}
\begin{aligned}
\mathbf{P^{R^\prime}} = S_z \mathbf{P^R}.
\end{aligned}
\end{equation*}

This keeps star camera framing consistent with what the user has defined on the main framing panel regardless of zoom level.

The simplest form is where the user can specify a zoom level for a script and use required points, $\mathbf{P^R}$, placed in guide camera image space as is. This gives the user creative freedom over actor framing at the edge and beyond of the star camera's frame. In this case, the points $\mathbf{P^R}$ are unchanged. Other settings allow for the zoom to automatically change depending on the size of the actor in frame.

\section{Results and Evaluation}\label{sec:Results}
The LookOut system has been used to film over $12$ hours of footage. To measure its strengths and find its weaknesses, we split up validation into five components: 

\begin{enumerate}
    \item Tracker performance, 
    \item Controller Evaluation,
    \item Hands-on evaluation by film-makers,
    \item Discussion of LookOut footage with senior film-makers, and \item Qualitative showcase of LookOut in different scenarios.
\end{enumerate}

For (1) and (2) we also compare performance against the DJI Osmo Mobile 3 in the supplemental. Note that illustrated footage in the supplemental website is slowed down to make ingesting telemetry data easier.

\begin{table*}[ht]
  \setlength{\tabcolsep}{3pt}
    \begin{tabular}{||c | c c c c c | c c c c c | c c c c c ||} 
     \hline
     & \multicolumn{5}{c|}{Market, 3m20s, one actor} & \multicolumn{5}{c|}{TwoPeople, 10m30s, two actors} & \multicolumn{5}{c|}{VOT-LT2019~\cite{VOT2019}, \textasciitilde2m24s, one target} \\ [0.5ex]
     \hline
     & $TP\uparrow$ & $MT\downarrow$ & $FP\downarrow$ & $\overline{D}\downarrow$ & T (ms)$\downarrow$ & $TP\uparrow$  & $MT\downarrow$ & $FP\downarrow$ & $\overline{D}\downarrow$ & T (ms)$\downarrow$ & $TP\uparrow$  & $MT\downarrow$ & $FP\downarrow$ & $\overline{D}\downarrow$ & T (ms)$\downarrow$\\ [0.5ex]
     \hline
Our Tracker & \textbf{4765} & 769 & \textbf{71} & \textbf{17.5} & 18.5  & \textbf{2655} & 562 & \textbf{132} & \textbf{34.8} & \underline{20.0}  & 0.200 & 0.764 & \textbf{0.036} & 65.0 & 11.7 \\
\hline
MOTDT~\cite{MOTDT} & \underline{4468} & 777 & \underline{362} & \underline{25.5} & 31.6  & 1770 & 1258 & 320 & \underline{42.1} & 32.3  & 0.229 & 0.690 & 0.081 & \underline{61.3} & 23.0  \\
\hline
SiamMask~\cite{siammask} & 4213 & \underline{78} & 1316 & 56.1 & \underline{14.9}  & \underline{1783} & \underline{72} & 1493 & 91.7 & 30.1  & \textbf{0.554} & \underline{0.153} & 0.292 & 85.5 & 12.8  \\
\hline
DaSiamRPN~\cite{dasiamrpn} & 2259 & 1732 & 1616 & 68.0 & \textbf{7.9} & 1709 & 528 & 1110 & 87.7 & \textbf{14.2}  & \underline{0.494} & 0.275 & 0.230 & 80.9 & 5.9  \\
\hline
DeepSORT~\cite{deepSORT} & 3961 & 554 & 1092 & 57.5 & 21.1  & 765 & 439 & 2143 & 147.0 & 22.6  & 0.186 & 0.752 & \underline{0.061} & 68.8 & 13.2  \\
\hline
KCF~\cite{kcf} & 623 & 3552 & 1432 & 94.2 & 87.3  & 377 & 2819 & \underline{151} & 123.3 & 73.7  & 0.173 & 0.736 & 0.090 & \textbf{47.2} & 4.4  \\
\hline
TLD~\cite{TLD} & 18 & \textbf{56} & 5533 & 239.4 & 32.2 & 670 & \textbf{1} & 2676 & 146.1 & 57.5  & 0.152 & \textbf{0.010} & 0.837 & 159.2 & 37.2  \\
\hline
\hline
SiamDW\text\_LT~\cite{SiamDW_2019_CVPR} & 4751 & 589 & 267 & 28.9 & 412.6 & 2693 & 175 & 479 & 42.7 & 1123.7 & - & - & - & - & - \\
\hline
    \end{tabular}
    \caption{We evaluate our tracker and other leading state-of-art real-time trackers on the VOT long-term tracking dataset. Other algorithms outperform ours on VOT. However, the VOT videos are qualitatively different in appearance from our use cases. So we introduce two further test sequences with 12,300 manually labeled annotations. These videos are more representative because of their cinematic style, both long and short term occlusions, and the presence of distractors, including people in cluttered environments. A high $TP$ (true positive) is obviously advantageous. A low $FP$ discourages the camera from moving onto a distractor. Some missed tracks, $MT$s, are tolerable, but especially after a long occlusion, missing the target could lead to catastrophic target loss. While a low $MT$ score is important, a trivial tracker that always outputs a bounding box, whether or not the target is occluded, would allow the tracker to be distracted. In the short term, this will lead to $FP$s, and in the long term, it will pollute that actor's appearance representation. $FP$s are especially detrimental for LookOut, because the camera is controlled by tracker output. An inaccurate position will move the camera away, further decreasing the chances of recovery and ruining a take. All run times include detector latency when appropriate. Detection based trackers are all run on tiny-YOLOv3 output. All trackers are run in a single thread, including ours. }
    \label{tab:trackerResults}
\end{table*}

\begin{figure}[ht]
  \centering
  \includegraphics[width=\linewidth]{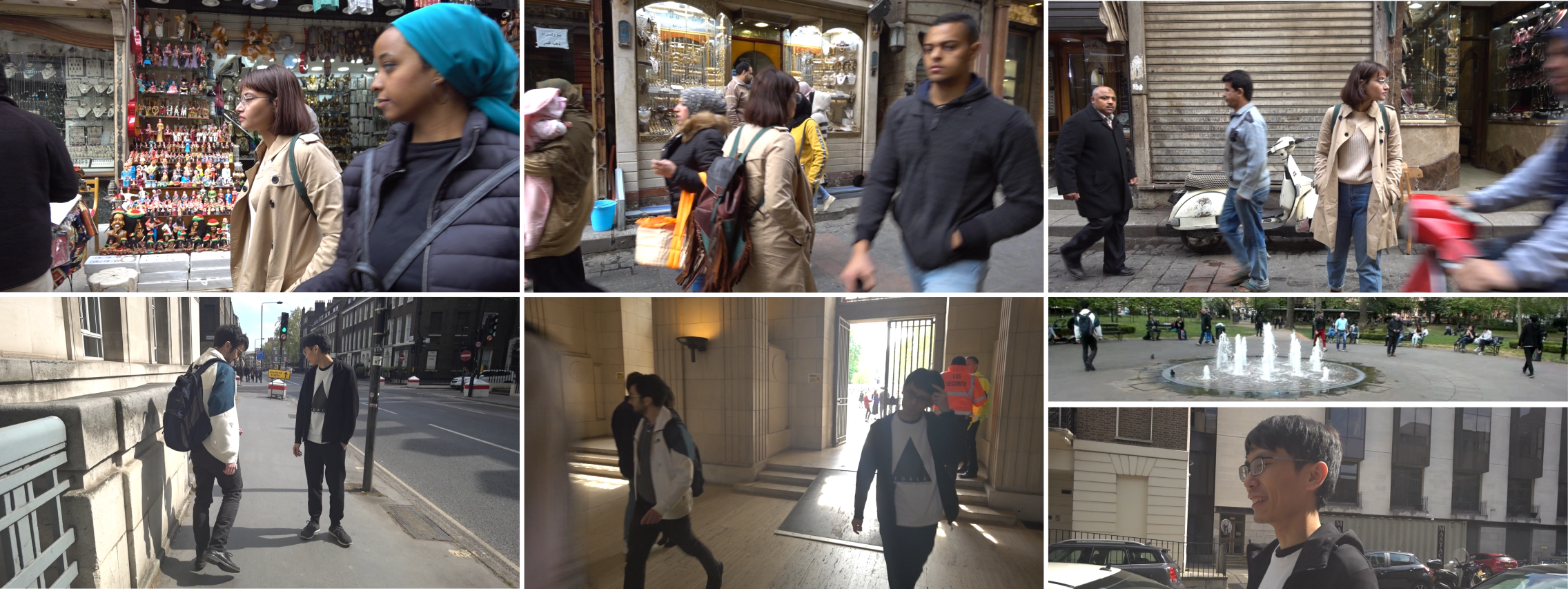}
  \caption{Sample frames from annotated videos used for benchmarks. Top: Market, a 3m20s scene of the actor in the beige coat walking through a crowded market. There are many occlusions in this scene, including where the target appears in frame with a different appearance than when they went into occlusion. Bottom: TwoPeople, a 10m30s scene of two actors on a walk through a campus and a park. Both actors wear similar looking clothes, occlude one another, disappear from frame entirely, are seen at different scales, and walk at various distances away from the camera.}
  \label{fig:market}
\end{figure}

\subsection{Tracker} \label{subsec:trackerResults}
We test our tracker's performance on the VOT Long-Term Challenge~\cite{VOT2019}, and on two long manually annotated videos that better represent our film-production use case. Market (one actor scene at 3m20s with annotations every frame) and TwoPeople (two actor scene at 10m30s with annotations every five frames) are challenging scenes with representative clutter, many occlusions by distractors, variable appearance before and after occlusion, and lighting changes (See Fig~\ref{fig:market}). Crucially, the subjects' appearance changes to something not seen before when emerging after an occlusion. While our tracker and others can sometimes be shown the subject from all angles to build a representative history, this test also checks for pickup-and-go filming performance, so no such five second grace training period is given. We ultimately advocate our tracker for the tracking of people in our use case. However, we include all videos from the VOT challenge in the comparison.

We describe in detail how these videos are annotated and the exact details of the associated metrics in the supplementary material. Broadly, a tracker is awarded a true positive ($TP$) point for a frame if it either correctly predicts the bounding box of the actor or correctly predicts that the actor is occluded. If a tracker outputs an incorrect bounding box, regardless of whether or not the actor is occluded, it is given a false positive point ($FP$) for that frame. If a tracker does not output a bounding box when the actor is not occluded, it is given a missed track ($MT$) point. We distinguish between $FP$ and $MT$ in this way to highlight errors that would point the camera away from the targets of interest, as is expressed with $FP$. We also compute the pixel distance between the center of the ground truth box and the center of the track, $D$, and obtain a mean over all updates, $\overline{D}$. We report raw unnormalised results for Market and TwoPeople (average of both actors) and normalized results on VOT-LT2019~\cite{VOT2019} sequences in Table~\ref{tab:trackerResults}.

\begin{table} 
  \setlength{\tabcolsep}{3pt}
    \begin{tabular}{||c | c c c c c ||} 
     \hline
     & $TP\uparrow$ & $MT\downarrow$ & $FP\downarrow$ & $\overline{D}\downarrow$ & T (ms)$\downarrow$\\ [0.5ex]
     \hline
Our Tracker & \textbf{0.822} & 0.152 & \textbf{0.026} & \textbf{22.6} & 19.3  \\
\hline
No Recovery & 0.745 & \textbf{0.088} & 0.166 & 35.8 & \textbf{18.7}  \\
\hline
Faulty Encodings & \underline{0.785} & \underline{0.140} & 0.075 & \underline{26.1} & 19.0 \\
\hline
Greedy Encodings & 0.698 & 0.234 & \underline{0.068} & 41.2 & 19.0  \\
\hline
Simple History & 0.688 & 0.182 & 0.131 & 50.7 & \underline{19.0}  \\
\hline
    \end{tabular}
    \caption{Ablation study of our tracker on the two test sequences and the metrics we establish in Section~\ref{subsec:trackerResults}. Simple history is a flavor of our tracker but with no feature history management, only the last seen $L$ encodings are stored in memory. \textit{No recovery} is our tracker but without a recovery stage. If a detection matches a target once, it is accepted as the target, leading to stray incorrect tracks on distractors, a high $FP$ score, and a lower $TP$ score in the long term. \textit{Greedy Encodings} stores a new incoming encoding into the feature gallery even if similar ones exist, filling up the gallery faster, thus leading to a restrictive appearance memory. \textit{Faulty Encodings} accepts detection encodings that are overlapped with other detections in the scene. This pollutes the gallery with noisy encodings and detracts from the tracker's ability to avoid distractors. Since the gallery sampling strategy is random, all trackers are run 40 times to ensure fairness.} \label{tab:trackerAblationResults}
\end{table}

We also ran a qualitative experiment with the leading VOT 2018 real-time tracker, DaSiamRPN~\cite{dasiamrpn}. We filmed an actor walking in a pedestrian area using both our tracker and DaSiamRPN~\cite{dasiamrpn} in separate takes. The rest of LookOut is kept constant, including actor weighting and actor specific leniency that both help to mitigate tracker noise and errors (but don't affect tracking). We run two takes each and show all takes in the supplementary video. While DaSiamRPN fails to track the actor in both takes, our tracker does. These takes show the importance of our robustness to imposters in filming.

\subsection{Controller Evaluation} \label{subsec:controllerEval}
In order to evaluate the controller components responsible for translating script commands to target camera frame radial velocities, we film multiple qualitative videos and also run an ablated version of the system. 

We film two takes of the same running scene at the same location and with the same predetermined path making sure to keep the relative motion between the camera and actor consistent. One take was filmed using our full system, including a minor leniency that's close to the minimum allowed. The second take was filmed with an ablated version of the control system, or `Standard Control.' The ablated version of the system passes raw tracker values as is to the PID controller without actor control weight adjustment (Sec.~\ref{subsec:transitions}) and the leniency mechanism (Sec.~\ref{subsec:leniency}). Fig.~\ref{fig:controllerAblation} shows camera frame radial velocities for both modes throughout this scene. Overall, the full controller satisfies scripted actor framing and largely ignores both actor track noise and camera translational motion that manifests itself as screen space motion. Following these internal and external noise sources would lead to an uncontrollably erratic camera as shown in the video and side by side radial velocities in  Fig.~\ref{fig:controllerAblation}. Please see this side-by-side comparison in the supplemental validation video and in the website as the video pair named \textit{Fully Ablated Control} under \textit{Control Ablation} for both the illustrated visualization and the star footage of this targeted A/B ablation comparison.

In the video \textit{Ablated Multi-Actor Weighting} under \textit{Control Ablation}, we show how using binary weights for actor script transitions produces a nervous erratic camera at best, and usually leads to a broken take. This happens because the error $T_c$ goes from being entirely Actor2 focused to entirely Actor1 focused, and vice versa, in one time step. This spikes the PID controllers leading to erratic corrections and a nervous camera. All other videos with multiple actors will show behavior with the method outlined in Sec.~\ref{subsec:transitions} and with leniency from Sec.~\ref{subsec:leniency}.

We also film scenes to show the effect of variable leniency on a single actor and for multiple actors, namely \textit{Hampstead Leniency Switch} and \textit{Clown and Calm} in the supplemental website. These videos demonstrate LookOut's ability to frame targets according to user defined leniency. In the supplemental website, please see other video illustrations of actor control weights, leniency ellipsis, and actor process variances displayed when available in filming metrics.

\subsection{Hands-on and End-to-end Evaluation}\label{Sec:HandsOnEval}
We designed and ran a small field study on an intermediate prototype, composed of two parts. Part 1 consisted of participants building a script using the LookOut GUI, while part 2 involved the same participants filming the scene they have programmed. 

\textbf{Participants}: In total we had 5 participants: four participants completed both parts, while one participant only completed part 1. We recruited the five volunteers (two female) by posting an advert on an amateur film-makers' group and through our own social networks. Three of them work within the film and entertainment industry (one lighting technician, one backstage support, and one director), while two are university students.

All participants had prior experience with filming, from beginner to amateur. Filming experience ranged from filming static scenes to action shots using Steadicams, from short clips for the Web to short movies.
None of the participants were familiar with computer vision, nor had they been exposed to the system before the study. One of the participants reported being familiar with Blockly from toys such as the Sphero\textsuperscript{TM}, which she previously encountered in her part-time work.

\textbf{Experimental Design:}
The study was designed to expose participants to the full operation of the system, from the creation of the configuration scripts using the GUI, to the actual filming of the action. To harmonize the task complexity across participants, we asked them to film a predefined sequence, communicated to them through a storyboard (printed in color on a single A3 page). Note that this form of study does not check creativity in run \& gun scenarios, but rather productivity~\cite{ShneidermanCreativity2007} when a DP is working solo.

Designing a suitable storyboard required careful consideration, to balance conflicting requirements. On one hand, we wanted a setting that really challenges visual tracking algorithms, and the storyboard to be particularly complex for a single operator to film in one shot. These requirements were to assess the system's ability to deal with challenging filming situations, and the ability of the GUI to expose a spectrum of behaviors. 

On the other hand, the storyboard design was constrained by concerns around the health and safety of participants (and to satisfy our research ethics review requirements). These concerns made us rule out any sequences involving stairs, streets with vehicles, or any other scenes that could be deemed unsafe. We also limited the number of actors required to two, and the overall study duration for each participant to one hour. 

After a number of iterations, involving consultation with a separate filmmaker, we agreed on the storyboard. Like many long takes, it incorporates a variety of shots, some of which would be quite hard to implement with standard filming techniques. One such difficult shot implements a sudden camera transition between the two actors, followed by the participant having to run to keep up with the actor named ``Blue.''

Another hard shot is the swooping pan where the camera starts low and ends up high as the participant moves around the tree until they are behind actor ``Red.'' This would normally be hard to execute as it involves the camera operator moving from a crouched to a standing position while ensuring the actor is kept within frame. With LookOut, the camera angle is adjusted automatically to frame the actor, letting the camera operator focus on their own movement. The storyboard can be found in the supplemental material.

As confirmation that the story and park setting were challenging themselves, two of our participants commented that, if they had the option, they would split the scene into separate shots (``I would segment the scene into different shots'' and ``normally I would split the scene into several parts'').

\begin{figure}[ht]
  \centering
  \includegraphics[width=0.9\linewidth]{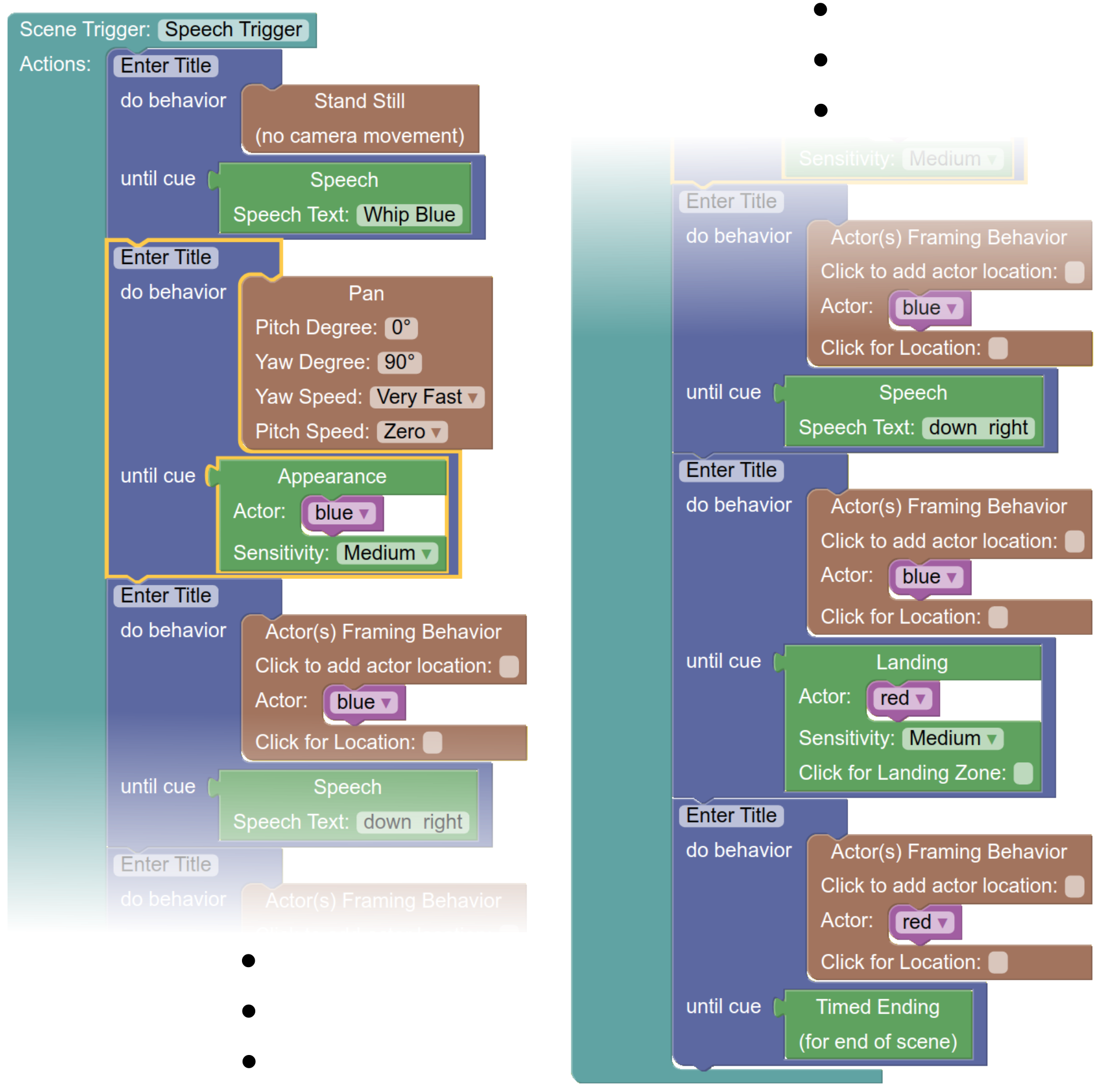}
  \caption{An example of a configuration script programmed by one of the participants. They opted to use whip pans and actor based cues to automate most of the camera's behavior change.}
  \label{fig:uiScriptFig}
\end{figure}

\textbf{Procedure:} Participants were given verbal instructions providing a brief overview of the user interface and the scene they were required to film. The setting was a local park, in late afternoon through dusk. Participants were handed a copy of the storyboard and left on a bench to create the required configuration scripts on a laptop running the GUI. Figure~\ref{fig:uiScriptFig} shows an example of a script created by a participant.

Once participants declared that they were satisfied with the scripts, they were provided with a quick overview of how the rest of LookOut works, and invited to start filming. As they tested their scripts, they were allowed to go back to the GUI and change aspects they thought did not work very well. For example, one participant went back and changed the speed of transitions, having realized that the ``very fast'' setting might miss locating the actor entirely.

Within 50 minutes of the start of the study, or as soon as participants filmed a scene they were satisfied with, the filming ended, and participants were asked to take part in a short interview (10 minutes) about their experience. 

\textbf{Configuration Scripts and GUI:}
All five participants who attempted part 1 of the study were able to successfully use the GUI to create configuration scripts to match the storyboard. This process lasted between 20 to 25 minutes, and was carried out independently by participants, although they were allowed to ask the experimenters questions. 
Participants were generally pleased with the UI's functionality. One participant commented that, ``programming the framing was like coding so it was simple enough'' while another participant stated that he was happy with the UI's behavior possibilities: ``already a lot with actor recognition and speech recognition.'' However, some participants did mention the need for a ``zoom function or focal length change.'' In addition, one participant wanted a feature to track objects: ``\eg if you wanted to track a statue while walking around it.'' Although LookOut supports object tracking, the UI did not offer this possibility at the time, only letting them select actors.

In some cases, after one or more attempts at filming the scene, participants realized that they were not happy with some of the details in their configuration scripts. In these cases, participants edited the configuration scripts using the GUI. 
In one case a participant realized that the duration for a timed cue was too short, so they adjusted the value. In another case they were not happy with the angle of the yaw in a pan, so they increased it. The adjustments took less than 5 minutes as the performed changes were minor parameter settings. No issues were reported or observed with the interface. 

These findings confirm that the task of scripting the behavior of the LookOut controller can be completed with minimal training by novice users. The editing of the parameters after a script was tested indicates that participants were able to relate the two, and could refine the script behavior to match their needs. 

\textbf{Filming and Resulting Footage:}
In the remaining 15-25 minutes, two of four participants had enough time to record a long take that they were happy with for this scene. In the other two cases, there were issues with the tracking of actors that led to the scene not being adequately filmed within the prescribed time frame. This was caused by the lighting being uncharacteristically bad: on most film sets, there would be procedures in place to reduce the effect of strong sunlight filtering through the trees, to keep the actors consistently lit.

One participant pointed out that even though they didn't have a view finder during filming, she could tell from the physical movement of the camera that it was smooth: ``from the physical movement of camera it looked smooth.''

Participants also spoke about the convenience of having an automatic movement of the camera as it meant they could focus on other aspects of the filming, such as keeping up with the actors. One participant described the task of keeping the camera focused on an actor as ``you can just track someone without caring about it.''

\textbf{Participant Comments:}
The aim of the storyboard was to have several different types of shots, some of them that would be harder to execute with traditional filming equipment. One of these shots involved having the camera quickly panning between the two actors: ``the whip pan was easier with the AI, it found and tracked the subject automatically. Otherwise I would have to rehearse that 3-4 times to get it correctly.'' By using LookOut, the participant was able to correctly capture the shot from the first take.

Participants were particularly pleased with using voice as a trigger for the next action in the scene: ``voice activating the cues worked very well.'' One participant stated that they ``could see directors using that to program in actor's lines.'' This feature simplified the filming process for participants, with all participants who attempted part two using speech triggers within their scripts.

When asked if there were other camera behaviors they'd like to see in LookOut, one participant mentioned tracking other objects, which we experimented with using a generic class object tracker (see ``Other Trackers'' on the supplemental website). Three participants mentioned zoom; although our current hardware limits optical zoom, we've made use of sensor based zoom (see ``Zoom''). The fifth participant said the existing behaviors were already a good toolbox, and specifically pointed out voice triggers as useful building blocks.

\textbf{Shot Breakdown:}
Takes were ruined either due to faulty tracking in bad lighting on the early version of the tracker used (46\%), participants forgetting to fire triggers they placed (12\%), voice recognition failure (8\%), and another 26\% miscellaneous (bystanders getting in the shot, actor mistakes, batteries running out, etc). The rest (12\%) yielded usable takes. The bulk of ruined takes come from tracker error, which motivated the development of our final proposed tracker and the underlying principles we lay out in Sec.~\ref{sec:trackerSection}. We have used the latest version of the proposed tracker for filming visually challenging scenes, including those in equally harsh lighting - ``Zoom Run'' and ``IRL Tracker Comparison''. 

\subsection{Critique by Senior Film-Makers}
\begin{figure}[ht]
  \centering
  \includegraphics[width=\linewidth]{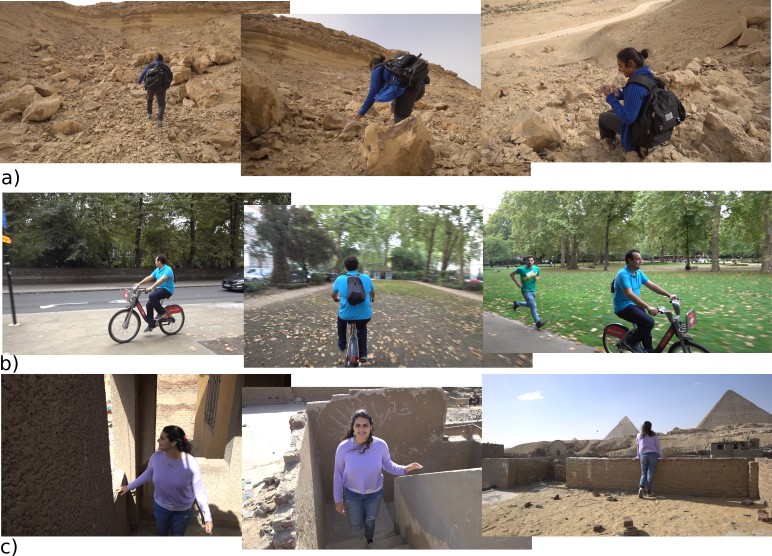}
  \caption{Videos shown to senior film-makers. a) Rocky escarpment - camera operator climbing on foot and with one hand free. a) Bike ride. Camera operator also riding a bike. c) Pyramids - camera operator walking backwards on stairs. }
  \label{fig:expperFeedbackVideo}
\end{figure}

We sought out three senior film-makers, separate from the film-makers who influenced the design of LookOut, and separate from those who did the Hands-On Evaluation (Section~\ref{Sec:HandsOnEval}). Each of them has been working as a professional Director of Photography, for $9$, $13$, and $25$ years respectively. Each of them has a mix of experience, in both scripted scenes with crew and actors, and run \& gun filming for documentaries or journalism. We interviewed them separately, each time showing the same three unedited video examples, shot using the LookOut system (see Fig~\ref{fig:expperFeedbackVideo}). We asked the same pre-defined set of questions to prompt them to think aloud while watching the videos.

The questions are listed in the Supplementary Material, but can be broadly grouped as concerning i) the equipment and people needed to film these long takes normally (without LookOut), and ii) critiques of both the footage and current LookOut capabilities.

First, to shoot such takes without LookOut, two of the film-makers have used drones, and would consider using them here, if a licensed pilot were available, and the noise wasn't prohibitive. Two of them said they would use cranes for video-A, if the budget allows. One complained, however, that multiple cranes have bad placement of viewfinders, resulting in them shooting blindly for long periods. For videos B and C, one said he would use a Steadicam, and the other two had specific two- or one-handed gimbals (like that modified for LookOut), that they would try again, despite having small and awkward viewfinders.

They each would need a second person at minimum, and usually more, to help with typical stabilization-only filming. Independently, they all said that if only one extra helper were available, then that person would be the spotter for the operator. A spotter physically guides the operator around obstacles.

Second, their views of the footage and the LookOut system were very positive, with some caveats. The two more senior ones expressed the sentiment that LookOut would have no place in a big-budget project, because the Director and DP can give orders verbally, that get carried out eventually. Also, those two would need to use LookOut multiple times before they'd trust its reliability, and ideally, prefer if colleagues make some films with it first. Transcribed interview quotes are in the supplemental material, and include comments such as ``That would be so helpful! Especially in those run \& gun situations, documentary, travel, journalism. If you're filming something that won't happen again, you can focus on the other things'' and ``I could be more creative once I got used to it.''

\subsection{Qualitative LookOut Results}
LookOut has been used by the authors, by test-subjects, and by novices who usually (but not exclusively) filmed using existing behavior scripts. A representative cross-section is shown in the supplemental videos web-page. Some noteworthy examples include sports where the operator is participating, such as skateboarding, or using one hand while \eg playing frisbee, scrambling, or cycling. For the Gnome and Plumbing-shop sequences, we filmed, as an exception, using the DaSiamRPN~\cite{dasiamrpn} tracker within LookOut, to cope with unusual object categories, though this required multiple takes. In contrast, the vast majority of takes using our tracker worked out on the first try.

\section{Limitations and Discussion}
The LookOut premise, software, and hardware, each have limitations. While it would be informative to do the end-to-end evaluation under run \& gun conditions, which represent the vast majority of users, those situations are rarely repeatable, and considered dangerous from an ethical experimentation perspective. That led us to use simple scripted scenarios for that evaluation. The senior film-makers are likely right that big-budget productions will be reticent to use LookOut. The field-study tested with participants from our low-budget demographic of film-makers with a fixed storyboard, but an ideal comprehensive user study would focus on adventure-athletes and journalists in somewhat dangerous conditions, to check real run \& gun scenarios.

\begin{figure}[h]
  \centering
  \includegraphics[width=\linewidth]{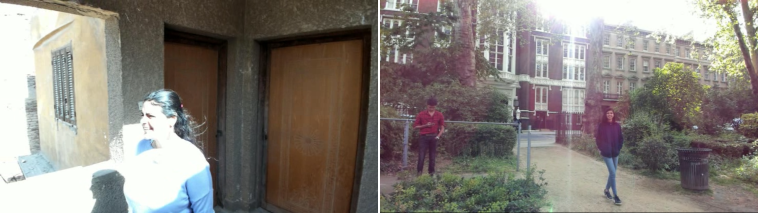}
  \caption{Harsh light and lens flares can upset the detector, and lead to gaps in tracking. If such a lighting change is fast enough and then long lasting, the tracker may not adequately associate new encodings with known actors, leading to a loss of tracking.}
  \label{fig:LimitationsLighting}
\end{figure}

The LookOut GUI worked better and more intuitively than expected. The detector and tracker combination too, perform admirably across really diverse scenarios, though they are designed initially for tracking actors across occlusions in hand-held films, and are unremarkable on the standard Computer Vision benchmarks MOT~\cite{MOT2016} and VOT~\cite{VOT2019}. The single weakest component across the LookOut system is the detector. We've seen it confuse the tracker when the actor hides or gets too small, there is too much motion blur, or actors wear the same uniform. For now, better detectors are available, but not with the low-latency required by the controller. LookOut is built in Python, which is not optimized for real-time and multiple threads. We chose this for easier comparison with other trackers and rapid prototyping, so efficiency gains are possible.
Like other appearance encodings, ours is sometimes susceptible to harsh and variable lighting (see Fig~\ref{fig:LimitationsLighting}), which makes the system most vulnerable at dusk or dawn, and possibly when switching between indoors and outdoors. On the fly camera image processing optimized to improve vision task performance similar to~\cite{TYY19} may help.

There are potentially two improvements for the hardware. First, some users requested that LookOut also manage focus-pulling and zooming, so this would require a star-camera where both focus and focal length is software-controllable in real-time. While we have showcased a hardware-limited version of zooming with our star-camera, further real-time control of both focus and focal-length is required to better realize this improvement. We have not found a suitable model yet.  Further, we use a guide camera with a limited field of view. $360^{\circ}$ cameras are rarely used for cinematic filming due to limited resolution, but could function as guide cameras. Then, new behaviors could better ``anticipate'' actors that aren't in-frame for the star camera yet. Extra sensing capability on the guide camera either through depth or infrared would further improve tracking and cinematic control. We will release the LookOut blueprints and downloadable system. 

\begin{acks}
We're grateful for thoughtful input of established filmmakers and videographers, including Luke Palmer, Gary Sinyor, Dave Ward, Tim Gardner, Tom Forsey, Jon Challicom, and Matt Judd. Thanks to all the user study participants and to our talented actors, including dancers Mark Turner and Diana Liasuk, and skateboarders Oliver Fine and Michael Jango'an. Through many development iterations, we are grateful for filming on-the-go, by Eric, Dan, Philipp, Pradyumna, Yushiang, David, Alex, Robert, Arjun, Emily, Ahmed, Habiba, Shahd, Asaad, Elbetity, Yomna, and Hala. Thanks to Peter Hedman and Graham Thomas for valuable discussions.
\end{acks}

\bibliographystyle{ACM-Reference-Format}
\bibliography{LookOut}


\newpage
\clearpage

\vskip .375in
\noindent{\huge \bf{Supplementary Material for \\
LookOut! Interactive Camera Gimbal Controller for Filming Long Takes \\
Videos at \textcolor{magenta}{\url{http://visual.cs.ucl.ac.uk/pubs/lookOut/scenes.html}}} \par}
\vspace*{24pt}

\vskip .5em
\vspace*{12pt}

\setcounter{equation}{0}
\setcounter{section}{0}
\setcounter{figure}{0}
\setcounter{table}{0}
\setcounter{page}{1}

\section{Interface for Filming Long Takes}
There is existing work in the literature that allows for expressing film scenes in a standardized form~\cite{wu2016analysing, psl, christianson1996declarative}. The Prose Storyboard Language~\cite{psl} provides a grammar that's readable by both humans and machines for expressing all aspects of a film scene  including where actors are located with respect to one another and the scene, camera framing and positioning, and how multiple shots are sequenced and stitched. The Prose Storyboard Language~\cite{psl} and other languages like it are powerful because they provide a standardized communication method throughout the entire film-making pipeline, ensuring the overall vision or goal of each scene and the overall film is met.

\begin{figure}[h]
  \centering
  \includegraphics[width=\linewidth]{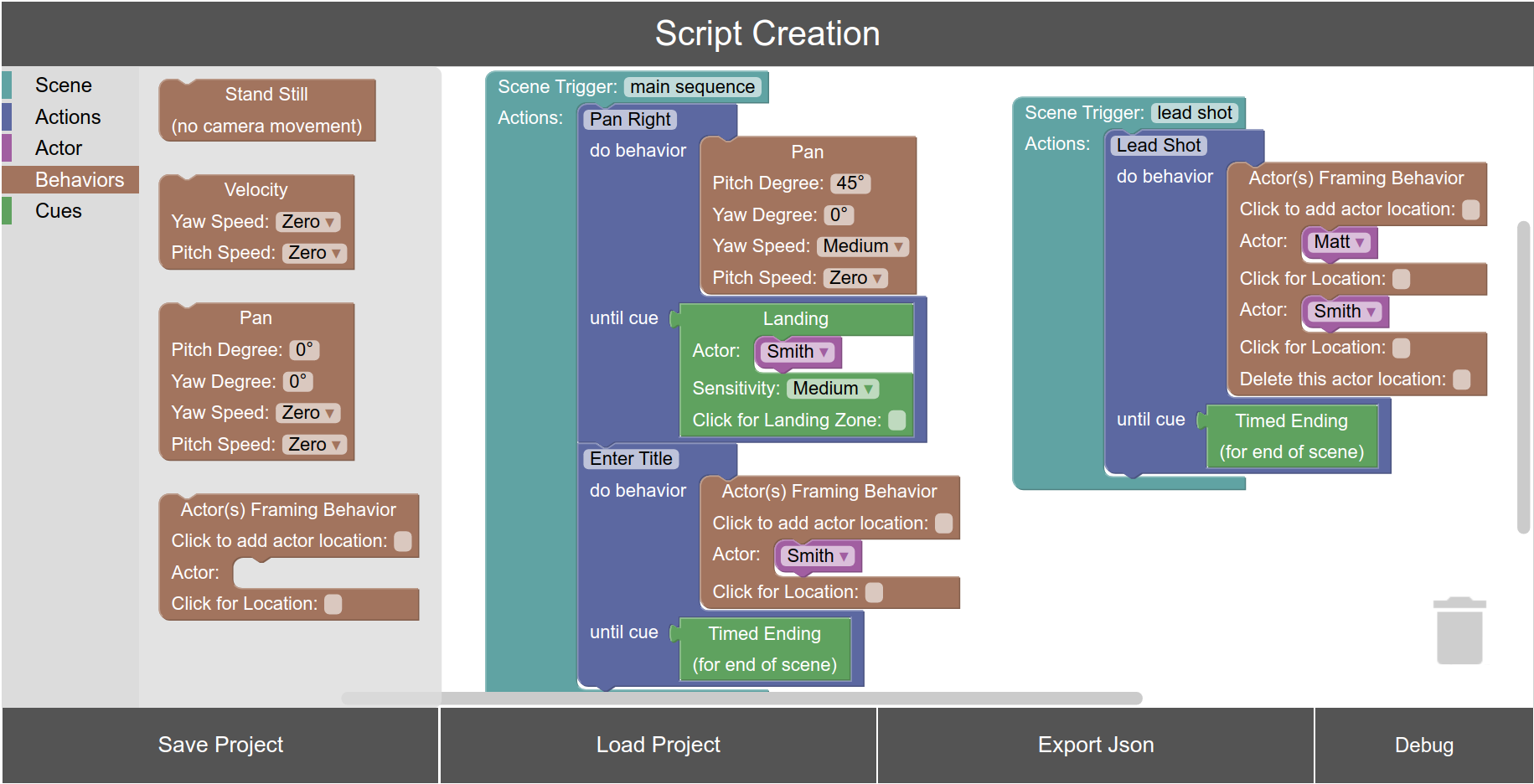}
  \caption{Workspace GUI for Script creation. Two scripts (shown with light blue ``blocks'') can be seen in this workspace. The left panel houses different structures for defining camera behavior. The Behaviors tab is open and displays some of the camera framing modes (in brown) available to the operator. Green ``blocks'' are cues, \ie events that are being monitored, to then conclude a behavior and/or start the next one.}
  \label{fig:ScriptWorkspace}
\end{figure}

Instead we focus on one key component of the pipeline, helping the camera operator film the shot. At the heart of LookOut's utility is the ability to offload the task of pointing the camera at actors from the camera operator. We call camera pointing like this a camera \textit{behavior} (Sec.~\ref{subsec:behaviors}). We house such behaviors in a \textit{script} (Sec.~\ref{subsec:scripts}). Each script is a series of sequential camera behaviors that are triggered in-turn when a \textit{cue} (Sec.~\ref{subsec:cues}) is hit. Cues can be as simple as a spoken word or as nuanced as an actor entering a particular part of the frame. The simplest script has no cues and only one behavior - keeping a single actor in frame for example - but they can be as complex as the operator wishes. The operator can also switch between multiple scripts during filming, allowing for flexibility depending on the circumstances. LookOut provides audible feedback through earphones to the operator, informing them when a cue is triggered or when a command is heard (Sec.~\ref{subsec:controlAndFeedBack}). Fig~\ref{fig:ScriptWorkspace} shows the GUI with an example script housing multiple behaviors and cues.

\subsection{Behaviors}\label{subsec:behaviors}
Within the LookOut GUI, the continuous domain of camera motions is organized into a menu of discrete and parameterized behaviors. Examples include standing still, or panning left $30^{\circ}$. The existing behaviors in LookOut come from requirements gathering with two film-makers, but further behaviors can be programmed in the future, and users can already cover a broad range of use cases by chaining behaviors together.

\textbf{Actor Based Framing:} In most filming scenarios, one or more humans is the focus of attention. So a shot is driven either by actor monolog/dialog, or by them performing physical movements. Unsurprisingly, a user designing a script with an actor-based behavior must first attach a specific ID to that behavior. An actor ID is mostly just a name for now, and the discriminative appearance info for each actor in the pool will only be filled in on-set at the startup phase. The user then specifies how this behavior frames that particular actor by placing a dot for the actor in the desired part of screen-space, \eg on the left of the frame (see Fig~\ref{fig:ActorsInGUI}). Often, scenes involve multiple actors. The interface for a multi-actor behavior is essentially the same, and LookOut will later optimize, striving to satisfy all the behavior's constraints. 

Not all movements the actors make on screen require the camera to move. Camera movements must be \textit{motivated}~\cite{katz2004cinematic}~\cite{brown2016cinematography}. A motivated camera movement or pan draws the attention of the viewer. Action shots may require tight and fast camera pans to keep a subject in frame, whereas a slow indoors shot would not benefit from quick camera pans when a subject rocks side to side.

For each actor's location, the user can specify an elliptical area of leniency, where actor movements are not immediately converted into camera movements. The LookOut GUI provides further behaviors for actor-framing. Especially for storyboarded takes, a path behavior lets the user spell out the framing of the actor over time. For example, a camera may pan to follow where an actor gazes when searching, or may pan to look ahead in a dynamic running shot. The path is constructed in screen space from a set of dots, with the distance between each point signifying how fast the camera moves in that part of the path.

\textbf{Non-Actor Behaviors:} A few of the available behaviors are independent of actor framing. A panning behavior makes the camera ``scan'' the scene, with a specified yaw or pitch direction, speed, and range. The banking behavior rolls the camera (in response to measurements from the IMU), simulating the effect of an aircraft dipping one wing while changing direction. The UI simply exposes the options for these behaviors using drop-down menus and text-entry fields. 

\subsection{Cues}\label{subsec:cues}
Camera movements are often deliberate, triggered by changes in the scene or the progression of the action. LookOut responds to these changes in the scene to initiate each successive behavior in the chain. To do this, LookOut monitors events that the user designated in the GUI as being significant cues for each context. LookOut informs the user through quick audio feedback when the next cue is hit. LookOut can currently monitor for the following cues:

\textbf{Actor Appearance and Disappearance:} Actors coming into and out of frame can signify a new camera behavior. This cue could signal that a new actor is to be followed or that an informative pan is to take place. The user can specify which actor LookOut should monitor, and how sensitive LookOut is to that change.

\begin{figure}[h]
  \centering
  \includegraphics[width=\linewidth]{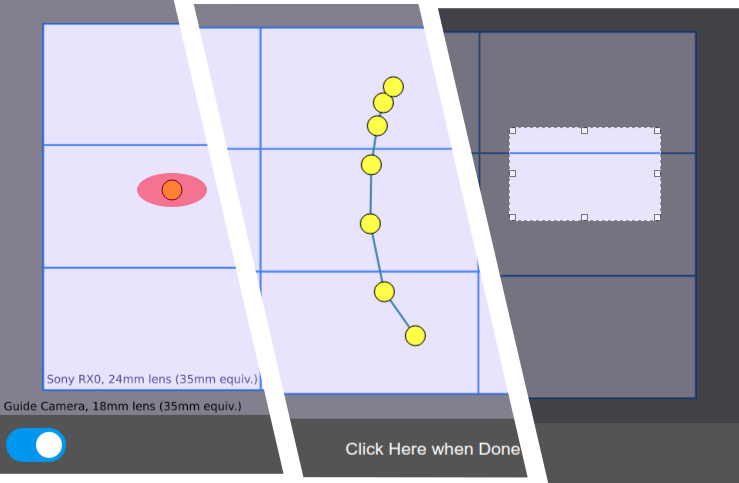}
  \caption{The operator can select where an actor must be positioned on screen with a yellow dot for location and red ellipse for leniency (left), string together multiple points for an actor path behavior (middle), and define an area for a landing zone cue (right). The blue grid represents the star camera's image space, including aspect ratio.}
  \label{fig:ActorsInGUI}
\end{figure}

\textbf{Landing Zone:} We adapt the concept of a landing zone into a cue. This cue is triggered when the requisite actor enters a specific user defined part of screen space (See Fig~\ref{fig:ActorsInGUI}).

\textbf{Elapsed Time:} Although rigid, we also allow control over how long a behavior runs, using an Elapsed Time cue. 

\textbf{Speech:} Speech recognition is included as a cue in LookOut, paired with speech synthesis. The aim is to give LookOut basic dialog-system capabilities, analogous to the verbal instructions between the director and a camera operator. The user types trigger words into the GUI when connecting a cue to a behavior, and either the operator or an actor with pre-defined lines wears a lapel microphone. The UI rejects trigger words that are too close to distinguish.


\textbf{Relative Actor Size:} This cue is useful for shots where the subject's relative image frame size is important to the narrative or action on frame. For example, a subject may appear from the distance, but the camera is to remain agnostic to them until they appear with a large enough screen size. 

\subsection{Scripts}\label{subsec:scripts}

A script is a preprogrammed sequence of camera behaviors. Ahead of filming, the user designs their long take using the LookOut GUI, typically on a laptop. They then export one or more scripts to the controller.

The chain of behaviors within a script is linked together by cues, which are explicit audible or visible events. The controller monitors for these events during filming, so a long take can be storyboarded and followed precisely, or it can be improvised in response to the operator's play-by-play instructions. Transitions between behaviors and between scripts are tuned to be responsive yet smooth. 

\subsection{Control and Feedback}\label{subsec:controlAndFeedBack}
 During filming, the system carries out the user's requests and provides audio feedback about which script is being used and which behavior the system is performing. These requests come in the form of speech commands, spoken by the user to interrupt a script, restart it, or jump to alternate scripts.

\textbf{Leniency From User Radii}
For advanced users, $\mathbf{q}$, $\mathbf{v}$, and $\mathbf{a}$ would be made available for fine control over leniency. However, in our UI implementation, leniency curves are abstracted into a single radii pair, $(r_x, r_y)$, for each axis that the user can specify via an ellipse in the UI. Note that this pair is normalized by guide camera image space size. We calculate each axis of the finer values using $\mathbf{r}$ as 
\begin{equation*}
\begin{aligned}
a &= 1.2r-0.005\text{,}\\
v &= 42 * (r-0.5)^{8}\text{, and}\\
q &= \frac{20}{r+0.01}\text{.}
\end{aligned}
\end{equation*}

\begin{center}
    
\end{center}

\section{Implementation Details} \label{sec:implemnentationDetails}

\subsection{Scripting and the GUI} 

\textbf{Implementation} 
We use the Blockly~\cite{blockly} library for constructing the GUI for stitching together behaviors and cues into scripts. The user selects a script file for LookOut to read at startup. We make HTML/JavaScript extensions outside Blockly for actor framing, path framing, and landing zone cue. The scripts are output as Json formatted files and read by the system at startup.

\textbf{Cues} 
With the exception of timed cues, all cues will continue waiting to be fired and LookOut defaults to the behavior preceding the respective cue meanwhile. When a cue is hit, the user receives audio feedback. If a cue isn't met, the user can push things along using an alternative mini-script they had programmed earlier.

\textbf{Actor Following Behavior} 
The present version of LookOut relies only on bounding boxes output from the tracker, so each actor's tracked point, $\mathbf{{p^{T}}}$, is set to halfway the width of the box in the x-axis and an offset below the top of the bounding box in the y-axis. This offset is set to 20\% of the vertical height of the bounding box. We've found that setting the y-value to the midway point between the min/max of the bounding box gives undesired framing when the target is close to the camera and the bounding box's bottom frame is clipped by the image boundaries. This can be improved with pose information.

\subsection{Voice Recognition} Script triggers can be fired anytime, but cue specific voice triggers can only be fired when relevant. For speed requirements, we require fast and reasonably accurate word or phrase recognition. This makes off-device services like Google's Cloud API unusable. Instead we make use of the Porcupine~\cite{porcupine} wake word detector.

\subsection{Camera Zoom} Our combination of hardware limits zoom control to three attainable distinct levels of sensor-based zoom - 1.0$\times$, 1.5$\times$, and 2.0$\times$. 

\begin{table}
  \setlength{\tabcolsep}{3pt}
    \begin{tabular}{||l l||} 
     \hline
     Operation & T (ms)$\downarrow$\\ [0.5ex] 
     \hline\hline
     Frame Grab & 8.2\\
     Frame Resize & 3.1\\
     Gimbal Control &  0.7\\
     Gimbal Metrics Retrieval & 8.9\\
     Miscellaneous & 1.6\\
     \textbf{Total} & 22.5\\
     \hline
     YOLOv3 and NMS & 13.1 \\
     Cosine Encoding Network & 3.8 \\
     \textbf{Total} & 16.9\\ [1ex] 
     \hline
     Tracker, Detection/Track assignment & 3.3 \\
     \hline
     \hline
    \end{tabular}
    \caption{Breakdown of task times in LookOut. There are three main threads. One main thread handles communication and control of the gimbal along with scripting logic. Gimbal Metrics Retrieval and Frame Grab are softly run in parallel. Tracker tasks are split into two threads, one handles GPU computation and the other handles the final detection/track assignment, and are run in a pipelined manner.}
    \label{tab:time}
\end{table}

\subsection{Latency and Threading} Table~\ref{tab:time} contains timing information for LookOut's main threads. Python's global interpreter lock prevents these threads from running truly in parallel, but there is a still a benefit to threading especially since some threads are blocked due to heavy I/O with either gimbal hardware or the GPU. The threads are:
\begin{enumerate}[label={\arabic*.}]
\item The main thread handles camera I/O, gimbal metrics and control, and main LookOut scripting logic. 
\item Tracker GPU tasks. Receives a camera frame from the main thread and passes detections and their appearance encodings to the CPU tracker thread.
\item Tracker CPU tasks. Receives detections and appearance encodings from the main thread, handles tracker logic, and sends results to the main thread.
\item Voice recognition logic, including Porcupine. Answers to main thread with recognized phrases.
\item Debug video storage and the debug display. Recieves current frame and control information from main thread.
\item Camera Zoom hardware interface. Answers to main thread with estimated zoom position and receives desired zoom levels.
\end{enumerate}

The overall latency from camera frame to tracker output and main LookOut thread refresh rate varies depending on a few factors, including camera exposure time, number of detectable objects in the scene, and the number of pending audio cues. 

\section{Tracker Details and Algorithm Pseudo-code} \label{sec:TrackerCode}

\subsection{Recovery Details}
In the main paper we outlined the recovery mechanism in the LookOut tracker. The length of recovery $R$, i.e. how many sequential frames a tracker must match to a detection before being counted as come out of recovery is variable. If the actor was only lost for a short period of time, typically a four tracker updates depending on system latency, then the $R$ is only two frames long. However if the track is lost for longer, the recovery length increases to four successful successive track hits. We do this to balance between being robust to distractors and accounting for short detector hiccups that would otherwise break tracking and require long recovery resulting in lost tracking. 

\subsection{Tracker Steady State Pseudo-Code}
See Algorithm~\ref{AlgTracker} for pseudo-code of the LookOut tracker's cost formulation strategy. The main explanation of the tracker is in the paper.

\begin{algorithm}[h]
\SetAlgoLined
\caption{\label{AlgTracker}Cost Matrix Formulation Pseudo-code for LookOut tracker.}
    \SetKwInOut{Input}{Input}
    \SetKwInOut{Output}{Output}
    \Input{A set of tracks $T = \{t_i, i \leq N\}$. A set of detections in the current frame, $D = \{d_j, j \leq M\}$. A chain of features for each track of length $L_i$, $F_i = \{f_k, k \leq L_i\}$, and a single feature associate with each detection $f_{d_j}$.}
    \Output{A matrix consisting of costs for each pair of detection, $d_j$, and track, $t_i$.}
    
    $C = [c_{ij}, i < N \text{ and } j < M]$ Overall cost matrix.\\ 
    $C^{\text{iou}} = [c_{ij}^{\text{iou}}, i < N \text{ and } j < M]$ IOU based cost matrix. \\
    $C^{\text{feature}} = [c_{ij}^{\text{feature}}, i < N \text{ and } j < M]$ Appearance based cost matrix.\\
    $C^{\text{avgfeature}} = [c_{ij}^{\text{feature}}, i < N \text{ and } j < M]$ Average appearance based cost matrix.
    
    \ForEach{track $t_i \in T$}{%
        $\text{overlapCount} = 0$ \;
        
        \ForEach{detection $d_j \in D$}{%
            $C^{\text{iou}}[i,j] = \text{computeIOU}(t_i, d_j)$\\
            $C^{\text{feature}}[i,j] = \text{computeAppearanceCost}(t_i, d_j)$\\
            $C^{\text{avgfeature}}[i,j] = \text{computeAvgAppearanceCost}(t_i, d_j)$
            
            \If{$c_{ij}^{\text{iou}} < \tau^{\text{overlap}}$} {%
                $\text{overlapCount} = \text{overlapCount} + 1$ 
            }
        }
        
        \If{$\text{overlapCount} < 2$ and $t_i$ is not lost or being recovered} {%
            \tcc{Only one detection is competing for this track spatially. Don't rely on appearance costs.}
            \ForEach{detection $d_j \in D$}{%
                $C^{\text{feature}}[i,j] = 0$\\
                $C^{\text{avgfeature}}[i,j] = 0$
            }
        }
    }
    $C = C_{iou} + C_{feature} + C_{avgfeature}$ 
    
\end{algorithm}

\subsection{Tracking Evaluation}
We evaluate trackers on two long representative scenes of our use case - Market (one actor scene at 3m20s with annotations every frame) and TwoPeople (two actor scene at 10m30s with annotations every five frames). 

We obtain a ground truth bounding box by manually annotating input videos at 740$\times$416. Market is annotated every frame, and TwoPeople has annotations for both actors every five frames. Similar to the VOT-LT challenge, we do not include estimated annotations for when the actor is completely occluded. Based on VOT~\cite{VOTmetrics} and MOT~\cite{MOT2016}, we employ metrics suited for actor tracking. All trackers are instantiated only once. Tracking-by-detection trackers are given the detection best fitting the groundtruth as a start point and other trackers are instantiated using the first groundtruth bounding box. Detection based trackers are all run on tiny-YOLOv3 output. Although trackers with multiple sequential components can be run in multiple threads for increased throughput, all trackers, including ours, are run in a single thread for fairness. 

We use a bounding box IOU threshold to determine if the correct bounding box is output. For matching tracker bounding-box output to the groundtruth target and distracters, we use an IOU of $0.5$. A tracker is awarded a true positive ($TP$) point for a frame if it either correctly predicts the bounding box of the actor or correctly predicts that the actor is occluded. If a tracker outputs an incorrect bounding box, regardless of whether or not the actor is occluded, it is given a false positive point ($FP$) for that frame. If a tracker does not output a bounding box when the actor is not occluded, it is given a missed track ($MT$) point. We distinguish between $FP$ and $MT$ in this way to highlight errors that would point the camera away from the targets of interest, as is expressed with $FP$. We also compute the pixel distance between the center of the ground truth box and the center of the track, $D$, and obtain a mean over all updates, $\overline{D}$. The center of frame is used instead of the tracker's output when the tracker is lost, and in case the tracker outputs some bounding box but the target is actually occluded.

Since our tracker has a random component, we average $40$ runs on the same LookOut backpack computer.

\section{Result Videos}

For videos, please see the supplemental website, \textcolor{magenta}{\url{http://visual.cs.ucl.ac.uk/pubs/lookOut/scenes.html}}, that contains guide camera footage with overlays of the LookOut system's inner processes. Every video also has an associated higher quality ``star" camera footage from a Sony RX0, albeit with a reduced bit-rate for consumption via web.  

The telemetry/illustrated footage is displayed at a reduced frame rate to make ingesting data easier by eye. Available telemetry data is listed on the video's page.

\section{Questions Asked of Senior Film-Makers}
\begin{figure}[ht]
  \centering
  \includegraphics[width=\linewidth]{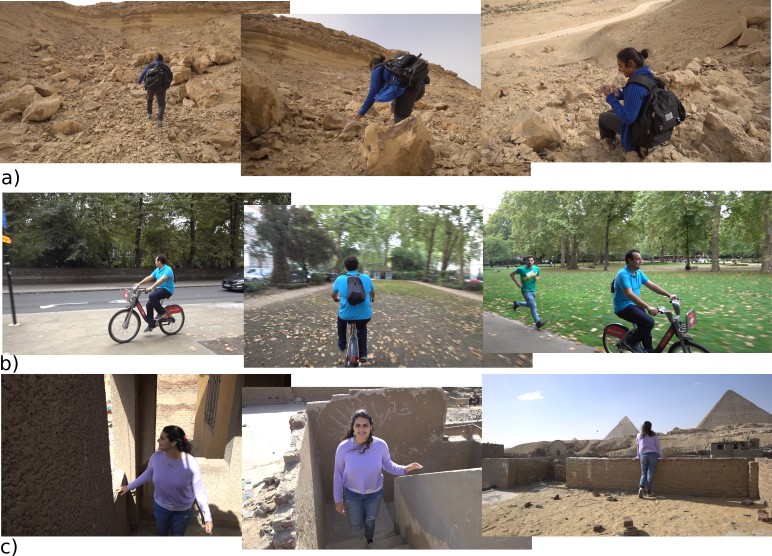}
  \caption{Videos shown to senior film-makers. a) Rocky escarpment - camera operator climbing on foot and with one hand free. a) Bike ride. Camera operator also riding a bike. c) Pyramids - camera operator walking backwards on stairs. }
  \label{fig:expertFeedbackVideo}
\end{figure}

These questions were put to our three most senior film-makers, while getting them to think-aloud while critiquing Videos A, B, and C in Figure~\ref{fig:expertFeedbackVideo}, all filmed using LookOut. These interviews and all experiments pre-date the Covid-19 pandemic.

\begin{itemize}
    \item How many people would you need to shoot a scene comparable to this?
    \item How many specialists would you need for it?
    \item What is the level of skill necessary for the operators to have for this type of shot? (1-5)
    \item What equipment would you need to shoot a scene comparable to this?
    \item Would you need to change the set? As in, bolt something to the floor, drill holes, dig, etc.
    \item Would you have a steadicam/gimbal operator do this or maybe have a crane?
    \item How much would it cost to get that equipment on site?
    \item How many takes / how consistent is the framing?
    \item How much planning goes into a shot like this?
\end{itemize}

To get natural reactions from the participants, we did not insist that they answer our specific questions. They did answer some of them. We made sure they at least gave their impressions about i) the equipment and people needed to film these long takes normally (without LookOut), and ii) critiques of both the footage and current LookOut capabilities.

Here are responses from each of the three film-makers, in turn. These responses were written in short-hand, and we omit various stories and deviations recounted during the interview, as had been agreed in advance, because some stories are unflattering, and we felt this could help make the responses more candid.

Quotes from film-maker ``W'':
\begin{itemize}
    \item To shoot a scene comparable to this? (video-A and overall): Depends on the budget. More people if you can afford it. Probably try to make due with 2. Myself and spotter, but still too rough a terrain and would need to rehearse a lot.
    \item (How many takes?) Just need a few more takes to find your feet and get it right; learn your movements and theirs.
    \item Skill? Probably prefer to get a crane operator, costs 2k per day. Similar price but less faff with a steadicam operator, but depends on the shot.
    \item I'd also think about using a drone on shots like these - just need someone who's licensed. But it's noisy, so no good if you're recording the dialog. Fine to try on a bigger production, and with no [low] wind. 
    \item (LookOut useful? Current capabilities?) Need to trust it first. Prefer if saw other film-makers using it repeatedly. Was the same for Red cameras. Saw the same with steadicam - nobody wants to be first, because an expensive set is expensive because of so many people and their time costs money.
    \item (How many specialists, skill level?) Hire someone on the basis of their experience or their style, not really quantifiable.
\end{itemize}

Quotes from film-maker ``F'':
\begin{itemize}
    \item To shoot a scene comparable to this? (video-A): myself + spotter, but still too rough a terrain. Need to rehearse. Could use a drone, but they're loud, a hassle, and then someone else is deciding.
    \item (video-A, How many takes?) Maybe just not attempt it, opt for simpler [shot], tripod.
    \item (video-B, special skill?): probably not [needed] for most people; common to be riding on something driven by others.
    \item (video-B would do differntly?) I like the orbiting, I'd like more face. Tighter framing of guy.
    \item (video-C): Gimbal, big monitor to make framing easier.
    \item (video-C how many takes?): Practice first, if they move too quickly, or you trip... Be CLEAR in your direction first. DJI Ronan - majestic mode can't respond to quick movements. 
    \item (Skill level? For all 3 videos:) If I really respected them, I'd tell them to get on with it. Vs. more novice, I'd give specific instructions. Not really detailed. Can always be a little different - so many variables. 
    \item (What equipment?) Xion crane - but motor on back covers LCD screen! And then mirorless cameras, screen is too small. Usually Sony A7 series, manual for Zoom, autofocus, or remote focus wheel in another hand. But those become a nuisance when limited time. (Later, after seeing LookOut and understanding it) Big thing with this, that you wouldn't need to worry about it. Use it when you can't look at the screen, and need to pay attention to other things. Low or high angles, when you can't see the viewfinder.
    \item (On seeing Lookout) That would be so helpful! Especially in those Run \& gun situations, documentary, travel, journalism. If you're filming something that won't happen again, you can focus on the other things.
    \item (Would you use it yourself?) Definite market for this; people could be funny about losing control [with motion control]; here purists could say it's part of the unique take. 
    \item (Asked us if they could try it out ``when we start selling them'' - really?) [Well, I ] can't see it in high-end commercial feature film. There, you have time on your side. ... Next time, let me know. I work with a lot of cinematographers. Lots of contacts who would like this. Should talk to DJI - probably most popular.
    \item (What's missing from LookOut?) Would you be able to control the zoom? 
\end{itemize}

Quotes from film-maker ``G'':
\begin{itemize}
    \item To shoot a scene comparable to this? (video-A): Use a gymbal camera with spotter to lead, move differently: smaller steps to minimize up and down; Or clear a path through boulders for walking, using a digger. Path needs to be out of shot obviously.
    \item (video-A equipment): Maybe 2-handed gymbal to reduce side-to-side; hold gymbal forward, at level of stomach.
    \item (video-B): Maybe Segway or cart driven by another person, so operator can focus on shooting.
    \item (Skill for video-B?) Should have understanding of camera work; need not be technical.
    \item (video-C): Definitely spotter leading me up the steps, on my shoulder, plus motorized single-handed gimbal, Easyrig [easyrig.se].
    \item (On seeing LookOut) That's amazing! 
    \item (LookOut useful?) Does it know something about the scene? (answered him and explained system) I probably would be comfortable with a speech command.
    \item (Would you use it yourself?) It wouldn't take me very long to trust it to get the shot I needed, if I got to see it working a few times. I could be more creative once I got used to it. [On big budget projects] not many things [shots] that I can only get the first time.
    \item (Current capabilities?) Focus is a massive consideration, can have it pulled just right, but dynamics of scene change. Or auto-focus goes wrong: change focus from person A to B, or from very near to very far, while I pan up to see a mountain. 
    \item (Other features?) Nothing right now. Is resistance pre-defined? Would like to adjust that, maybe dynamically: example, walk toward building and then look up, so need resistance. Sony FS7 with Ronin S, film a lot in slow motion. Start at 50 or 100fps then slow down.
\end{itemize}
\section{User Study}
\begin{figure*}[h]
  \centering
  \includegraphics[width=0.8\linewidth]{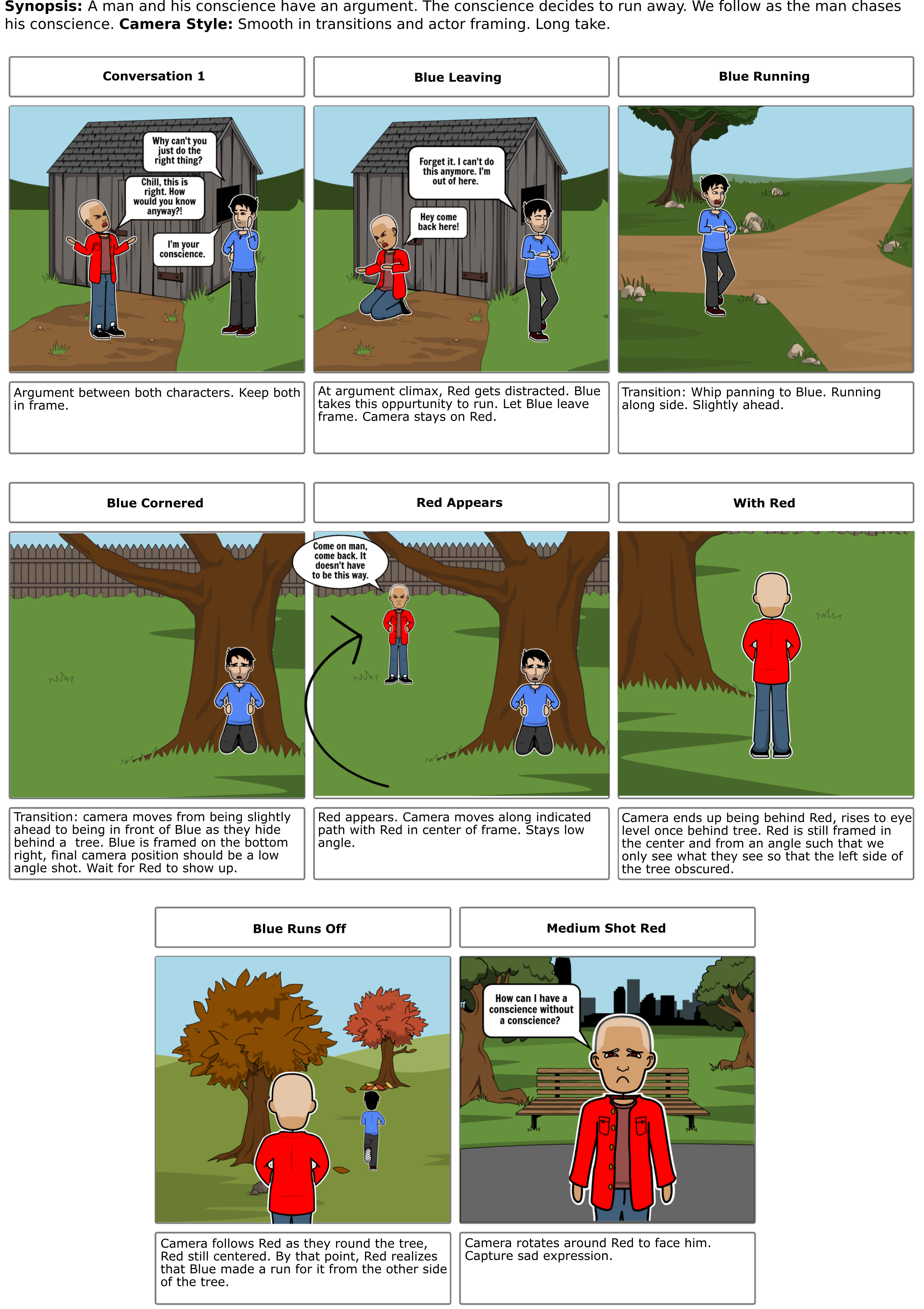}
  \caption{Storyboard given to participants to film using LookOut during the user study.}
  \label{fig:storyboardFig}
\end{figure*}

The storyboard given to participants during the user study is in Fig.~\ref{fig:storyboardFig}.
\end{document}